\documentclass[aps,prb,twocolumn,eqsecnum,superscriptaddress,showpacs,10pt]{revtex4-1}
\usepackage{natbib}
\usepackage{mathbbol,amsmath,amsfonts,amssymb,bm,color,dsfont,bbold}
\usepackage{graphicx,psfrag,array,braket}
\usepackage{float}
\usepackage[T1]{fontenc}
\usepackage{orcidlink}
\usepackage{soul}
\allowdisplaybreaks

\providecommand{\ignore}[1]{}
\providecommand{\aucmnt}[1]{#1}
\def\be{\begin{equation}}
\def\ee{\end{equation}}

\renewcommand{\aucmnt}[1]{}
\newcommand{\Comment}[1]{}
\newcommand{\Eq}[1]{Eq.~(\ref{#1})}

\begin{document}
\title{From frustration-free parent Hamiltonians to off-diagonal long-range order: Moore-Read and related states in second quantization}
\author{Fanmao Zhang}\affiliation{College of Physics and Electronic Science, Hubei Normal University, Huangshi 435002, China}
\author{Matheus Schossler\,\orcidlink{0000-0002-4681-059X}}\affiliation{Department of Physics, Washington University, St.
Louis, Missouri 63130, USA}
\author{Alexander Seidel\,\orcidlink{0000-0002-7008-7063}}
\affiliation{Technical University of Munich, TUM School of Natural Sciences, Physics Department, 85748 Garching, Germany}
\affiliation{Munich Center for Quantum Science and Technology (MCQST), Schellingstr. 4, 80799 M{\"u}nchen, Germany}
\affiliation{Department of Physics, Washington University, St.
Louis, Missouri 63130, USA}
\author{Li Chen\,\orcidlink{0000-0002-4870-9065}}
\affiliation{College of Physics and Electronic Science, Hubei Normal University, Huangshi 435002, China}
\date{\today}
\begin{abstract}
We construct a recursive second-quantized formula for  Moore-Read Pfaffian states. We demonstrate the utility of such second-quantized presentations by directly proving the existence of frustration-free parent Hamiltonians, without appealing to polynomial clustering properties.
Furthermore, we show how this formalism is connected to the existence of a non-local order parameter for Moore-Read states and give a proof that the latter exhibit off-diagonal long-range order (ODLRO) in these quantities. We also develop a similar second-quantized presentation for the fermionic anti- and PH-Pfaffian states, as well as $f$- and higher wave paired composite fermion states, and discuss ODLRO in most cases.
\end{abstract}
\maketitle

\section{Introduction}

The past few decades have witnessed tremendous efforts in the study of strongly correlated systems, including unconventional superconductors\cite{doi:10.1146/annurev-conmatphys-020911-125055,doi:10.1146/annurev-conmatphys-031113-133822,RevModPhys.87.855}, quantum spin liquids\cite{Savary2016,RevModPhys.88.041002,RevModPhys.89.025003,Senthilscience},  as well as fractional quantum Hall(FQH) systems\cite{RevModPhys.71.S298,RevModPhys.75.1101,compositeReview,QHCFTRevMod}. In FQH systems, the kinetic energies of electrons are quenched as electrons occupy a certain Landau level, rendering Coulomb interactions as the major term in the Hamiltonian.  Closed form for the ground state of the many-body Coulomb interaction is difficult to obtain; thus, theorists resort to model Hamiltonians for which the prototypical trial state of the closed-form wave function is the \textit{exact} unique densest zero mode (zero-energy ground state with the minimum total angular momentum).  These model Hamiltonians include two-body pseudopotential\cite{haldane_hierarchy} for Laughlin state\cite{Laughlin}, two-body parent Hamiltonian\cite{PhysRevLett.124.196803} for \textit{unprojected} Jain composite fermion state\cite{jainbook,compositeReview}, three-body parent Hamiltonian\cite{GREITER1992577,GreiterPf} for Moore-Read Pfaffian state\cite{MR},  and general multi-body parent Hamiltonians for Read-Rezayi states\cite{RR,Simonetal,Simonetal2}. Of all FQH states, much attention has been paid to those with non-Abelian anyonic excitation,  a key necessary ingredient for the topological quantum computation\cite{RevModPhys.80.1083,KITAEV20032}. A typical example is Moore-Read Pfaffian state, which is constructed from correlators in conformal field theory\cite{CFT}.

In the study of  model FQH states and their corresponding parent Hamiltonian, it is common practice to  focus  on the \textit{first-quantized} wave functions, whose algebraic clustering properties when two or more particles come together are traditionally utilized to construct closely related  first-quantized parent Hamiltonians.
More recently, a second-quantized approach has been developed to yield alternative, second-quantized presentations of FQH models states, study their parent Hamiltonians, and establish new such Hamiltonians\cite{Chen14,Chen19,PhysRevLett.124.196803}.
In particular, this approach has proven effective in constructing parent Hamiltonians\cite{PhysRevLett.124.196803} for unprojected Jain composite fermion states,
which are, in general, not fully characterized by conventional clustering properties.
It has also been used to explain the existence of a frustration-free parent Hamiltonian as a consequence of the matrix product structure of the Laughlin state\cite{Seidel21}.
Furthermore, it inspired a picture for
particle fractionalization\cite{2021exact} that largely recovers a symmetry between quasiholes and \textit{quasiparticles}, which is typically obscure in traditional treatments.
A strength of the second-quantized approach is that it allows rigorous statements about the zero mode space of some frustration-free solvable models where traditional methods
are inadequate. This is particularly so in the context of parton-like states (see Refs. \onlinecite{2bar11,wu2017new,Parton_antiPfaffian,1bar2bar111,fwave,PhysRevB.103.085303,PhysRevResearch.3.033087,PhysRevB.103.155103,Balram:2021opn,PhysRevB.105.L241403} and references therein), where Landau-level mixing leads to wave functions
that are no longer represented by holomorphic polynomials, barring established techniques from being used to prove uniqueness and/or completeness of
zero mode trial wave functions.
Alternative methods to achieve such statements have recently been developed, emphasizing largely second-quantized methods over first-quantized ones.
In some cases, one can develop the entire theory surrounding certain classes of trial wave functions,  their parent Hamiltonians, and their associated zero mode spaces
using exclusively second-quantized formalism that nowhere references the polynomials associated to first-quantized wave functions.
This has, in particular, been done for Laughlin states\cite{Chen14} as well as all composite fermion states in the positive Jain sequence\cite{PhysRevLett.124.196803}.
Here, the construction of traditional polynomial trial wave functions is replaced by certain recursion relations in particle number that allow
the second-quantized trial states to be created from the vacuum via a corresponding operator product.
The prototypical version of such products is Read's presentation \cite{ReadOP} of the Laughlin state as a ``condensate'' involving
a non-local order parameter (which was originally given in a mixed first/second-quantized notation).
Analogously, second-quantized constructions were recently discussed for composite fermion states\cite{Chen19}.

In this paper, we put forth similar developments that yield a fully second-quantized construction of the Moore-Read sequence,
and a concurrent discussion of its parent Hamiltonians. Our main result is a fully second-quantized expression of Moore-Read states
as an operator product acting on the vacuum.
As is well-known, the parent Hamiltonians of Moore-Read states involve three-body terms\cite{GREITER1992577,GreiterPf}.
While higher-body terms are quite common in the literature of quantum Hall parent Hamiltonians
\cite{PhysRevB.65.245309,PhysRevB.80.121302,PhysRevB.87.155426,PhysRevB.87.245129,PhysRevB.87.245425,PhysRevB.98.201101},
the discussion is typically limited to the lowest Landau level utilizing first quantization. While we will not
leave the lowest Landau level in this paper, one byproduct of our approach will be the extension of second-quantized methods
so far exclusively applied to two-body interactions to solvable models involving higher-body terms. We thus make manifest how the ``frustration-free property'' of the Moore-Read state and its parent Hamiltonian arises in second quantization. We will further utilize these results to demonstrate the existence of off-diagonal long-range order in Moore-Read states. Finally, we will extend several of these results to the anti-Pfaffian and PH-Pfaffian states.

This paper is organized as follows.
In Sec.~\ref{MRPfaffianandPH}, we set up the problem.
In Sec.~\ref{fermionre}, we postulate a  second-quantized recursive formula~\eqref{refor1} for fermionic (bosonic) $\nu=1/M$ Pfaffian state, whose zero mode property is proven in Sects.~\ref{zm2} and~\ref{zmprpty}.
In Sec.~\ref{ff}, we perform a root analysis of the recursively defined state.
In Sec.~\ref{fracc}, we  obtain its second-quantized non-local order parameter and prove the existence of off-diagonal long-range order.
In Sec.~\ref{HigherPfaffian}, we generalize to the Pfaffian states with higher angular momentum pairing. 
In Sec.~\ref{aP}, we obtain the second-quantized recursive formulas~\eqref{anrec} and~\eqref{PHrec} for fermionic anti- and PH-Pfaffian states, based on the recursive formula for fermionic Pfaffian state. 
We present discussion and outlook in Sec.~\ref{DO}.

\section{Second-quantized  Moore-Read Pfaffian state\label{MR2ndquant}}
\subsection{Moore-Read Pfaffian state and its parent Hamiltonian}\label{MRPfaffianandPH}

In this section, we review some defining properties of the Moore-Read state and its parent Hamiltonian, and establish the second-quantized formulation of these properties.

The parent Hamiltonian for  the $\nu=1/M$ fermionic (bosonic) Moore-Read Pfaffian state\cite{MR}, whose first-quantized wave function is  given by
\be\label{1stPfaff}\text{Pf}\left(\frac{1}{z_i-z_j}\right)\prod\limits_{k<l}\left(z_k-z_l\right)^M\ee  with even (odd) positive integer $M$ for fermions (bosons) respectively,
consists of two-body and three-body projection operators\cite{RRcount},
\be\label{totalH} H=H^{(2\text{bd})}+H^{(3\text{bd})}.\ee

The two-body projection operator $H^{(2\text{bd})}$ in second quantization is of the following form\cite{ortiz},

\be\label{2bodyPH} H^{(2\text{bd})}=\sum_{\substack{0  \leqslant m<M-1\\ (-1)^m=(-1)^{M-1}}}\sum\limits_{J\in \mathbb Z^{0+}} T^{(2\text{bd},m)\dag}_J  T^{(2\text{bd},m)}_J,\ee
where the positive-semidefinite two-body fermionic (bosonic)  operator $T^{(2\text{bd},m)\dag}_J  T^{(2\text{bd},m)}_J$
is the second-quantized form of the Haldane
$V_m$ pseudo-potential\cite{haldane_hierarchy}.
That is, it
projects onto an antisymmetric (symmetric)   two-body state of relative angular momentum $m\hbar$ and total angular momentum $J\hbar$ in the lowest Landau level (LLL). In disk geometry, it can be given a concrete form via
\begin{align} T^{(2\text{bd},m)}_J=2^{\frac{1-J}{2}}\sum_{k} &\, p_{m,\frac{J}{2}}(k)\sqrt{{J \choose m}{J \choose \frac{J}{2}+k}} \notag\\ & \times c_{\frac{J}{2}-k}c_{\frac{J}{2}+k},\end{align}
and similar expressions hold in other geometries\cite{ortiz}.
Here, ${J\choose m}=J!/(J-m)!m!$ is the binomial coefficient, $ c_{i} $ is a fermionic (bosonic) operator that annihilates a particle of angular momentum $i \hbar $ in the LLL. Throughout this paper, we are dealing with LLL orbitals on the disk, so only those $ c_{i} $ with nonnegative $i$ are of concern to us. We, therefore, let $ c_{i}=0 $ whenever we formally encounter negative $i$ in the calculation.
$p_{m,\frac{J}{2}}(k)$ is a polynomial in $k$ of degree $m$ and parity $(-1)^m$, whose expression is given by
\be\label{2poly}
\begin{split}
p_{m,\frac{J}{2}}(k)= &(-1)^{m+\frac{J}{2}-k}\frac{{m \choose J/2-k}}{{J \choose J/2-k}}\\&\times {_2} F_1 (-\frac{J}{2}+k,-J+m,1-\frac{J}{2}+k+m,-1)
\end{split}
\ee
with ${_2}F_1$ the hypergeometric function.

The zero mode, or ground space of $H^{(2\text{bd})}$
is spanned by the $\nu=1/(M-1)$ Laughlin state and its zero-energy excitations, which physically represent the edge- and quasihole-excitations of this state.
The zero-mode condition associated with $H^{(2\text{bd})}$ can be cast as
\be\label{zo2} T^{(2\text{bd},m)}_J \ket{\psi_{\text{zero}}}=0\ee
for all $J$ and $m$ in \Eq{2bodyPH}. This zero-mode condition is clearly invariant under the formation of new linearly independent {{linear}} combinations of the operators $T^{(2\text{bd},m)}_J$, and thus can be written as
\be\label{zo3} Q^{(2\text{bd},m)}_J \ket{\psi_{\text{zero}}}=0\ee
in terms of simpler operators
\be Q^{(2\text{bd},m)}_J=\sum_{\substack{0 \leqslant i_1,i_2 \leqslant J\\ i_1+i_2=J}}\frac{(i_1-i_2)^m }{\sqrt{i_1!i_2!}} c_{i_2}c_{i_1}.\ee
Here, $J$ and $m$ run over the same values as before. The simple monomial form of the last expression offers yet a more condensed version of the two-body zero-mode condition. Defining the operators
\be\label{QP} Q^{(2\text{bd},{\cal P})}_J=\sum_{\substack{0 \leqslant i_1,i_2 \leqslant J\\ i_1+i_2=J}}\frac{{\cal P}(i_1,i_2) }{\sqrt{i_1!i_2!}} c_{i_2}c_{i_1},\ee
where $\cal P$ is any polynomial in two variables of the requisite symmetry, we may equivalently cast
\Eq{zo3} as
\be\label{zo4} Q^{(2\text{bd},{\cal P})}_J \ket{\psi_{\text{zero}}}=0,\ee
where $J$ runs over all nonnegative integers as before, and $\cal P$ can be {\em any} polynomial of degree less than $M-1$. To see the equivalence with
\Eq{zo3}, write $\cal P$ in terms of variables $i_1+i_2$ and $i_1-i_2$, and note that $i_1+i_2$ is a constant in the definition of \Eq{QP}.

We will now similarly cast the zero-mode condition associated with $H^{(3\text{bd})}$. $H^{(3\text{bd})}$, as given in the literature \cite{RRcount, Simonetal2},
is a three-body projection operator that projects onto states of relative angular momentum $3M-3$.
To make the claim even stronger, we also include the three-body projection operator that projects onto states of relative angular momentum $3M-2$. 
The Moore-Read state will be the unique zero mode of the resulting Hamiltonian within its angular momentum sector with or without the addition of the $3M-2$ term.
Note, however, that the latter must be taken to vanish identically if $M=2$(fermionic case) or $M=1$(bosonic case), since the corresponding three-body states do not exist\cite{Simonetal2}.
The second-quantized form for $H^{(3\text{bd})}$ is thus given by
\be\label{pro} H^{(3\text{bd})}=\sum\limits_{t=3M-3}^{3M-2}\sum\limits_{J\in \mathbb Z^{0+}} T^{(3\text{bd},t)\dag}_J T^{(3\text{bd},t)}_J,\ee
with

\be\label{qR}\begin{split}
T^{(3\text{bd},t)}_J=\sum_{\substack{0 \leqslant i_1,i_2,i_3 \leqslant J\\ i_1+i_2+i_3=J}}\frac{{\cal Q}_t(i_1,i_2,i_3)}{\sqrt{i_1!i_2!i_3!}} c_{i_3}c_{i_2}c_{i_1}.
\end{split} \ee
Here, $t$ runs over an index set that labels an orthonormal basis of three-particle states with total angular momentum $J$ and relative angular momentum $t$ (all in units of $\hbar$).
Any such state can be expressed via \Eq{qR} through an appropriately chosen polynomial ${\cal Q}_t$ in three variables, of the requisite symmetry for fermions/bosons. (${\cal Q}_t$ will also depend on $J$ and $M$, we will, however, leave this understood.)
${\cal Q}_t$ can be chosen to be of degree $t$ (not necessarily homogeneous).

The zero-mode condition
associated to $H^{(3\text{bd})}$ then reads,
in complete analogy with the two-body case,
\be\label{zo} T^{(3\text{bd},t)}_J \ket{\psi_{\text{zero}}}=0\ee
for all $J\geqslant 0$ and $t=3M-3,3M-2$.

For general $M$, the polynomials ${\cal Q}_t$ are rather complex, even more so than their two-body counterparts~\eqref{2poly}.
Luckily, we will not need their precise form.
For similar reasons, though perhaps less well known, the zero-mode condition~\eqref{zo} can be given an equivalent form analogous to \Eq{zo4}. To this end, we define generic three-body destruction operators

\be\label{q3}\begin{split}
Q^{(3\text{bd},{\cal Q})}_J=\sum_{\substack{0 \leqslant i_1,i_2,i_3 \leqslant J\\ i_1+i_2+i_3=J}}\frac{{\cal Q}(i_1,i_2,i_3)}{\sqrt{i_1!i_2!i_3!}} c_{i_3}c_{i_2}c_{i_1}
\end{split} \ee
with $\cal Q$ a polynomial in three variables and of the desired (anti-)symmetry.
By definition, the zero modes we are interested in satisfy {\em both} the two-body and the three-body zero-mode conditions \Eq{zo4} and \Eq{zo}. In this case, we may, however, replace the three-body zero-mode condition~\eqref{zo} with the seemingly stronger condition
\be\label{zo5} Q^{(3\text{bd},{\cal Q})}_J \ket{\psi_{\text{zero}}}=0\ee
for all integers $J\geqslant 0$ {\em and} all three-variable polynomials $\cal Q$ of degree less than or equal to $3M-2$. Clearly, ensuring \Eq{zo5} is sufficient to ensure that \Eq{zo} is also satisfied. Below we will show that our second-quantized expression for the Moore-Read state satisfies both Eqs.~\eqref{zo4} and~\eqref{zo5}. It is thus, in particular,  a zero mode of the Hamiltonian \Eq{totalH}. In turn, any state that is a zero mode of this Hamiltonian, and has the same angular momentum as the Moore-Read state, must be equal to the Moore-Read state~\eqref{1stPfaff} itself (up to a constant). This follows from known spectral properties of this Hamiltonian\cite{RRcount, Simonetal2}. We will thus be able to establish, without referring to any explicit first-quantized polynomial construction, that the second-quantized expression, which will constitute the main result of this work below, is the Moore-Read state.

It may be instructive, however, to understand why
fulfillment of the stronger equation~\eqref{zo5} by the Moore-Read state is not coincidental, but indeed
a zero mode satisfying both \Eq{zo4} (or any of its equivalents) and \Eq{zo} also
satisfies \Eq{zo5}.
This may be done as follows.
One may convince oneself that any three-particle state generated from the vacuum $\ket{0}$ via $(Q_J^{3\text{bd},{\cal Q}})^\dagger \ket{0}$, with $\cal Q$ of degree $L$, lies in the subspace of relative angular momentum less than or equal to $L$.
(Conversely, if a three-particle state of given total angular momentum $J$ has relative angular momentum $L$, it can be written in this way by a polynomial of degree $L$.)
Hence, for $L=3M-2$ and at given $J$ these
three-particles states span the subspace spanned by the states associated with the ${\cal Q}_t$ defined after \Eq{qR} {\em and} (all) additional states of relative angular momentum {\em less than} $3M-2$. However, it is well known that zero modes of $H^{(2\text{bd})}$ in \Eq{2bodyPH} are automatically annihilated by three-particle projection operators onto states with relative angular momentum less than $3M-3$. It is for this reason that such three-particle projection operators are usually excluded from \Eq{totalH}. Hence, in the presence of the two-body constraint~\eqref{zo4}, the three-body constraint~\eqref{zo5} becomes truly equivalent to that of~\eqref{zo}.

\subsection{Recursive formula for the fermionic (bosonic) Pfaffian state}\label{fermionre}

With its essential defining properties now in place, we postulate the following
 second-quantized recursive formula for the Moore-Read ``Pfaffian'' state, whose first-quantized wave function is  \Eq{1stPfaff}:

 \begin{subequations}\label{refor}
 \begin{align}\label{refor1}
\ket{\text{Pf}_{N+2}}=& \frac{1}{N+2}\sum\limits_{l=0}^{M-1}(-1)^l {M-1 \choose l} \sum\limits_{r,k=0}^{MN+M-1 }\sqrt{r!k!}\notag\\&\times  c_{r}^\dagger c_{k}^\dagger S_{MN+M-1-l-r}S_{MN+l-k}\ket{\text{Pf}_{N}}\end{align}  for even nonnegative particle number $N$, where the particle-number-conserving operator $S_\ell$ is defined in \Eq{S} below.
The beginning of the recursion is defined by $\ket{\text{Pf}_{0}}=\ket{0}$. 
As we comment below, the recursion \eqref{refor1} can be viewed as a purely second-quantized version of a
``mixed'' first/second-quantized presentation of the Pfaffian state that has already appeared in the original work by Moore and Read\cite{MR}. While \Eq{refor1} can be derived directly from the Moore-Read wave function \eqref{1stPfaff},
we will emphasize here that one does not need to make contact with this first-quantized wave function, nor any other presentation given originally by Moore and Read, in order to show directly that \eqref{refor1} defines the densest zero mode of a frustration-free parent Hamiltonian. Our approach is thus intrinsically second-quantized.

In addition, the recursion \eqref{refor1} generalizes a similar second-quantized recursion for the Laughlin state\cite{Chen14} that, in turn, can be seen to be a (purely) second-quantized rendition of Read's presentation\cite{ReadOP} of the Laughlin state as ``Bose condensate'' of certain (non-local) ``order parameter'' operators that are off-diagonal in particle number. An important distinction between \Eq{refor1} for the Moore-Read state and the earlier recursions for the Laughlin state is that
we are increasing particle number by two, reflecting the paired nature of the state. However, the
Moore-Read state with odd particle number can also be accessed in this framework, simply via removal of one particle from $\ket{\text{Pf}_N}$ with  even $N$. We will comment in detail on particle removal further below.
Wherever desired, we will notationally condense \Eq{refor1} to
 \begin{align}\label{refor2}
\ket{\text{Pf}_{N+2}}={\cal R}_N\ket{\text{Pf}_{N}}\,,
\end{align}
\end{subequations}
where ${\cal R}_N$ denotes the operator on the right hand side of \Eq{refor1}.

To prove \Eq{refor}, we will utilize the strategy set up in the preceding section. That is, we will establish the state $\ket{\text{Pf}_N}$ as defined in \Eq{refor1} to be a zero mode of the parent Hamiltonian~\eqref{totalH}, which uniquely defines the state given that it has the proper total angular momentum. This serves the important additional goal of exposing the inner workings that render complex (long-ranged) second-quantized positive semi-definite Hamiltonians -- like the one in question -- frustration free. It is also for this reason that we proceed without making any essential use of the first-quantized wave function~\eqref{1stPfaff}. We will, however,  comment on how \Eq{refor1} {\em could} be derived in the first-quantized manner in  Appendix~\ref{1stproof}.

To proceed, we make contact with operator formalism first established in Ref. \onlinecite{ortiz, Chen14, Mazaheri14}, and then generalized to composite fermions in multiple Landau levels in Ref. \onlinecite{Chen19}.
The  $S$ operator\cite{Chen14,Chen19} in \Eq{refor1}, which originates from $\prod_{i<j}\left(z_i-z_j\right)^M$ in the first quantization,  is defined as \begin{align}\label{S}
& {   S_\ell } = {( - 1)^\ell }\sum\limits_{n_1 + n_2+\cdots+n_M= \ell }  e_{n_1} e_{n_2}\cdots e_{n_M}\quad \text{for} \quad \ell\geqslant 0,\notag\\&
   S_\ell=0\quad \text{for} \quad \ell<0,\end{align}
where $e_n$, in turn, is the particle-number-conserving operator that, in first quantization, multiplies the wave function with the elementary symmetric polynomial
$2^{-n/2}\sum_{1\leqslant i_1<i_2\cdots<i_n\leqslant N} z_{i_1}z_{i_2}\cdots z_{i_n}$. Second-quantized representations of these operators and other generators of symmetric polynomials have been discussed in detail in Ref. \onlinecite{Mazaheri14}. We have
\begin{align}\label{e}
e_n =& \frac{1}{n!} \sum_{l_1,\dots ,l_n =  0}^{ + \infty } \sqrt {l_1 + 1} \,c_{l_1+1}^\dag\sqrt {l_2 + 1} \,c_{l_2+ 1}^\dag \cdots\sqrt {l_n+ 1}\notag\\&\times \,c_{l_n + 1}^\dag    c_{ l_n} \cdots c_{ l_2}c_{ l_1}\quad \text{for}~n>0,\notag\\
e_0=&\mathbb{1},\notag\\ e_n=& 0 \quad \text{for}~ n<0.\end{align}
This then allows for recursive generation of the second-quantized Moore-Read state via \Eq{refor} and \Eq{S}.

$e_n$ is related to power-sum symmetric polynomial operator \be\label{pows} P_d = \sum_{r=0}^{+\infty} \sqrt{\frac{(r+d)!}{r!}}\,c^\dagger_{r+d}c_{r}\ee for $d  \geqslant 0$ by Newton-Girard relation\cite{Mazaheri14,Chen19}, \be\label{ng}e_n=\frac{1}{n}\sum\limits_{d=1}^{n}(-1)^{d-1} P_d \,e_{n-d}. \ee
The action of $P_d$ on an $N$-particle state is that of multiplying its first-quantized wave function with the power-sum symmetric polynomial
$P_d \equiv 2^{-d/2}\sum_{i=1}^N z_i^d$.

$P_d$ is a ``zero mode  generator'' in the sense that when acting on a zero mode $\ket{\psi_{\text{zero}}}$, as defined by Eqs.~\eqref{zo4} and~\eqref{zo5},
it gives a new zero mode. The reason is that $Q^{(2\text{bd},\cal P)}_J P_d\ket{\psi_{\text{zero}}}=0$ since $[Q^{(2\text{bd},\cal P)}_J,P_d] $  is of the form $Q^{(2\text{bd},\cal P')}_{J-d}$, with $\cal P'$ a polynomial of degree no larger than that of $\cal P$.
Thus, $[Q^{(2\text{bd},\cal P)}_J,P_d]$ vanishes on zero modes, by \Eq{zo4}.
For analogous reasons, we also have $Q^{(3\text{bd},\cal Q)}_J P_d\ket{\psi_{\text{zero}}}=0$. By Newton-Girard formula, every $e_n$ can be  expressed in terms of all $P_d$ with $d=1,2,\dots n$. Therefore,  $e_n$ and $S_\ell$ are also zero mode generators.

Another important property of  $S_\ell$ is that different $S_\ell$ commute with each other. The commutative property of  $S_\ell$, can likewise be established by first establishing the commutativity of the $P_d$ amongst themselves, and then extending
this property to the $e_n$ via Newton-Girard relations.

A centerpiece of this work and the machinery to follow
is the description of the effect of the removal of a single particle in state $r$ from the state $\ket{\text{Pf}_{N+2}}$ in terms of the addition of a particle to the state $\ket{\text{Pf}_{N}}$, plus operators generating a ``correlation hole'' just big enough such that the net effect is the local charge depletion described by $c_r$:
\begin{align}\label{re1}
c_r \ket{\text{Pf}_{N+2}}=&\frac{\sqrt{r!}}{2}\sum\limits_{l=0}^{M-1}(-1)^l {M-1 \choose l} \sum\limits_{k=0}^{MN+M-1 }\sqrt{k!}\, c_{k}^\dagger\notag\\&\times   [S_{MN+M-1-l-r}S_{MN+l-k}+(-1)^{M-1}\notag \\ & \times  S_{MN+M-1-l-k}S_{MN+l-r}]\ket{\text{Pf}_{N}}.\end{align}
This equation can actually be derived as a pure consequence of \Eq{refor1}, that is, without resorting to the first-quantized wave function of the Moore-Read state. We show this in the Supplemental Material\cite{Supplement}.
The derivation is lengthy, however. To the less patient reader, we thus offer an alternative proof of \Eq{re1} (and by extension \Eq{refor1}) that uses the first-quantized wave function. This proof is given in Appendix~\ref{1stproof}.

We shall now proceed to show that the recursion \Eq{refor1} defines the Moore-Read state at filling $1/M$ by showing that it is a zero mode of the appropriate parent Hamiltonian at the proper angular momentum. We begin with the two-body terms.

\subsection{Proof that recursively defined Pfaffian state is a zero mode of the two-body Hamiltonian~\eqref{2bodyPH}}\label{zm2}

We prove by the method of mathematical  induction that the state as recursively defined in \Eq{refor1}
is a zero mode of all $Q^{(2\text{bd},\cal P)}_J$ with degree of $\cal P$ less than $M-1$,  thus a zero mode of the two-body Hamiltonian~\eqref{2bodyPH}.

{\textit{Proof.}} We begin the induction by proving the claimed property directly for $\ket{\text{Pf}_{0}}=\ket{0}$,  $\ket{\text{Pf}_{2}}$ and $\ket{\text{Pf}_{4}}$.
By using the recursive formula \Eq{refor1}, the second-quantized form of $\ket{\text{Pf}_{2}}$ is
\begin{align}\label{psi2root}
\ket{\text{Pf}_{2}}=&  \frac{1}{2}\sum\limits_{l=0}^{M-1}(-1)^l {M-1 \choose l} \sum\limits_{r,k=0}^{M-1 }\sqrt{r!k!} \, c_{r}^\dagger c_{k}^\dagger  \notag\\&\times S_{M-1-l-r}S_{l-k}\ket{0}\notag\\=&  \frac{1}{2}\sum\limits_{l=0}^{M-1}(-1)^l {M-1 \choose l} \sqrt{(M-1-l)!l!}\notag\\&\times c_{M-1-l}^\dagger c_{l}^\dagger \ket{0}.\end{align}
In the above calculation of $\ket{\text{Pf}_{2}}$, we have used the fact that  $S$ operator is the sum of products of $e$ operators, which have annihilation operators on the right, thus $S_{M-1-l-r}S_{l-k}\ket{0}$ vanishes unless $M-1-l-r=0$ and $l-k=0$. The second-quantized form of $\ket{\text{Pf}_{4}}$ is given in \Eq{psi_4}.

$\ket{\text{Pf}_{0}}$,  $\ket{\text{Pf}_{2}}$ and $\ket{\text{Pf}_{4}}$ are annihilated by all  $Q^{(2\text{bd},\cal P)}_J$ with degree of $\cal P$ less than $M-1$,  since $\ket{\text{Pf}_{0}}$ is vacuum and
\be\begin{split} Q^{(2\text{bd},\cal P)}_J\ket{\text{Pf}_{2}}=&(-1)^{M-1}\delta_{J,M-1}\sum_{l=0}^{M-1}(-1)^{l}{M-1 \choose l}\notag\\&\times {\cal P}(l,M-1-l)=0,\end{split} \ee due to \Eq{combi}.
The proof that $\ket{\text{Pf}_{4}}$ is annihilated by all  $Q^{(2\text{bd},\cal P)}_J$ is given in Appendix~\ref{2annihilate}.

Now we
establish the induction step, assuming that
\begin{align}  Q^{(2\text{bd},\cal P)}_J\ket{\text{Pf}_{N}}=0\end{align}
holds for all $J$, with some $N \geqslant  4$. Then we have
\begin{widetext}\begin{align}\label{Q2psi}
& (N+2)Q^{(2\text{bd},\cal P)}_J\ket{\text{Pf}_{N+2}} \notag\\ =& Q^{(2\text{bd},\cal P)}_J\sum\limits_{l=0}^{M-1}(-1)^l {M-1 \choose l} \sum\limits_{r,k=0}^{MN+M-1 }\sqrt{r!k!}\, c_{r}^\dagger c_{k}^\dagger  S_{MN+M-1-l-r}S_{MN+l-k}\ket{\text{Pf}_{N}} \notag\\
=& 2 \sum_{\substack{0  \leqslant  i_1,i_2 \leqslant J\\ i_1+i_2=J}}\frac{{\cal P}(i_1,i_2)}{\sqrt{i_2!}}c_{i_2}\sum\limits_{l=0}^{M-1}(-1)^l {M-1 \choose l} \sum\limits_{k=0}^{MN+M-1 }\sqrt{k!}\, c_{k}^\dagger  \notag \\ & \times [S_{MN+M-1-l-i_1}S_{MN+l-k}+(-1)^{M-1}  S_{MN+M-1-l-k}S_{MN+l-i_1}]\ket{\text{Pf}_{N}}\notag\\
& +2 (-1)^M\sum_{\substack{0  \leqslant  i_1,i_2 \leqslant J\\ i_1+i_2=J}} {\cal P}(i_1,i_2)\sum\limits_{l=0}^{M-1}(-1)^l {M-1 \choose l}S_{MN+M-1-l-i_2}S_{MN+l-i_1}\ket{\text{Pf}_{N}}\notag\\
&+ \sum\limits_{l=0}^{M-1}(-1)^l {M-1 \choose l} \sum\limits_{r,k=0}^{MN+M-1 }\sqrt{r!k!}\, c_{r}^\dagger c_{k}^\dagger  Q^{(2\text{bd},\cal P)}_J S_{MN+M-1-l-r}S_{MN+l-k}\ket{\text{Pf}_{N}}\notag\\
=& 4 Q^{(2\text{bd},\cal P)}_J\ket{\text{Pf}_{N+2}} +2 (-1)^M\sum_{\substack{0  \leqslant  i_1,i_2 \leqslant J\\ i_1+i_2=J}} {\cal P}(i_1,i_2)\sum\limits_{l=0}^{M-1}(-1)^l {M-1 \choose l}S_{MN+M-1-l-i_2}S_{MN+l-i_1}\ket{\text{Pf}_{N}},
\end{align}\end{widetext}
where we have used \Eq{refor1} in the first step, $c_{i_2}c_{i_1}c_{r}^\dagger c_{k}^\dagger =\delta_{r,i_1}c_{i_2}c_{k}^\dagger+(-1)^{M-1}\delta_{r,i_2}c_{i_1}c_{k}^\dagger+\delta_{k,i_2}c_{i_1}c_{r}^\dagger+(-1)^{M-1}\delta_{k,i_1}c_{i_2}c_{r}^\dagger-\delta_{r,i_1}\delta_{k,i_2}+(-1)^{M}\delta_{r,i_2}\delta_{k,i_1}+c_{r}^\dagger c_{k}^\dagger c_{i_2}c_{i_1}$ in the second step, and \Eq{re1} in the last step so as to re-assemble the first expression after the second step into the first expression on the last line. We have also used the identity $Q^{(2\text{bd},\cal P)}_J S_{MN+M-1-l-r}S_{MN+l-k}\ket{\text{Pf}_{N}}=0$ since $S_{MN+M-1-l-r}$ and $S_{MN+l-k}$ are zero mode generators, and $\ket{\text{Pf}_{N}}$ is assumed to be a zero mode.

Now we need to simplify the last term
\begin{align}\label{last2} &\sum_{\substack{0  \leqslant  i_1,i_2 \leqslant J\\ i_1+i_2=J}} {\cal P}(i_1,i_2)\sum\limits_{l=0}^{M-1}(-1)^l {M-1 \choose l}\notag\\&\times S_{MN+M-1-l-i_2}S_{MN+l-i_1}\ket{\text{Pf}_{N}}.\end{align}

Under  change of variables, $i_1-l=i'_1$ and $i_2+l=i'_2$, the above term becomes
\begin{align}&\sum\limits_{l=0}^{M-1}(-1)^l {M-1 \choose l}\sum_{\substack{-l \leqslant  i'_1 \leqslant J-l  \\ l \leqslant  i'_2 \leqslant J+l\\ i'_1+i'_2=J}}{\cal P}(i'_1+l,i'_2-l) \notag\\&\times  S_{MN+M-1-i'_2}S_{MN-i'_1}\ket{\text{Pf}_{N}}.\end{align}

Now we shall use an important identity, \be\label{constraint1} S_\ell \ket{\text{Pf}_{N}}=0  \quad\text{for  $\ell>MN$.}\ee The reason for its validity is the following: the state $\ket{\text{Pf}_{N}}$ has $N$ particles, while nonzero $S_\ell$ is defined as ${( - 1)^\ell }\sum\nolimits_{n_1 + n_2+\cdots+n_M= \ell }  e_{n_1} e_{n_2}\cdots e_{n_M}$, in which $e_{n_i}$ will move the orbitals of $n_i$ particles for $i=1,2,\cdots,M$. For  $\ell>MN$, there must be an $n_i$  larger than  the particle number $N$, thus $S_\ell$ annihilates  $\ket{\text{Pf}_{N}}$ in this case.

As a result of the above identity, the lower limit of both $i'_1$ and $i'_2$ can be changed to 0, which does not affect the summation. Therefore, the upper limit of  both $i'_1$ and $i'_2$ can be changed to $J$ on account of $i'_1+i'_2=J$.  After the change of limits of summations, \Eq{last2} can be finally simplified to
\begin{align}& \sum_{\substack{0  \leqslant  i'_1,i'_2 \leqslant J\\ i'_1+i'_2=J}}\bigg[\sum\limits_{l=0}^{M-1}(-1)^l {M-1 \choose l}{\cal P}(i'_1+l,i'_2-l) \bigg] \notag\\&\times S_{MN+M-1-i'_2}S_{MN-i'_1}\ket{\text{Pf}_{N}}, \end{align} which vanishes since the summation in the square bracket is exactly 0 as a result of \Eq{combi},  considering that the degree of $\cal P$ is less than $M-1$.

After this lengthy simplification, we obtain $(N+2)Q^{(2\text{bd},\cal P)}_J\ket{\text{Pf}_{N+2}}=4Q^{(2\text{bd},\cal P)}_J\ket{\text{Pf}_{N+2}}$.
Therefore, if $\ket{\text{Pf}_{N}}$ is a zero mode of all  $Q^{(2\text{bd},\cal P)}_J$ for some $N  \geqslant  4 $, so will be $\ket{\text{Pf}_{N+2}}$. By mathematical induction, the fermionic (bosonic) Pfaffian state, as recursively defined in \Eq{refor1},
is thus a zero mode of the two-body Hamiltonian~\eqref{2bodyPH}.\hfill $\blacksquare$

\subsection{Proof that recursively defined Pfaffian state is a zero mode of the three-body Hamiltonian~\eqref{pro}}\label{zmprpty}

Next, we prove by the method of mathematical induction that the fermionic (bosonic) Pfaffian state,  as recursively defined in \Eq{refor1},
is a zero mode of all $Q^{(3\text{bd},\cal Q)}_J$ with degree of $\cal Q$ less than $3M-1$, thus a zero mode of the three-body Hamiltonian~\eqref{pro}.

{\textit{Proof.}} To begin the induction,
we prove the claimed property directly for $\ket{\text{Pf}_{0}}=\ket{0}$,  $\ket{\text{Pf}_{2}}$, $\ket{\text{Pf}_{4}}$ and $\ket{\text{Pf}_{6}}$.

It is easy to see that $\ket{\text{Pf}_{0}}$ and $\ket{\text{Pf}_{2}}$ are  annihilated by all $Q^{(3\text{bd},\cal Q)}_J$, since in these cases the particle numbers are less than three. We prove that $\ket{\text{Pf}_{4}}$  and $\ket{\text{Pf}_{6}}$ are annihilated by all $Q^{(3\text{bd},\cal Q)}_J$ in Appendix~\ref{3annihilate}.

Now prove the induction step and assume that
\be Q^{(3\text{bd},\cal Q)}_J\ket{\text{Pf}_{N}}=0 \quad\text{for all $J$ and some $N \geqslant  6$.}\ee Similar to \Eq{Q2psi}, we obtain
\begin{widetext}
\begin{align}
& (N+2)Q^{(3\text{bd},\cal Q)}_J\ket{\text{Pf}_{N+2}} \notag\\
=& Q^{(3\text{bd},\cal Q)}_J\sum\limits_{l=0}^{M-1}(-1)^l {M-1 \choose l} \sum\limits_{r,k=0}^{MN+M-1 }\sqrt{r!k!}\, c_{r}^\dagger c_{k}^\dagger  S_{MN+M-1-l-r}S_{MN+l-k}\ket{\text{Pf}_{N}} \notag\\
 =& 3 \sum_{\substack{0  \leqslant  i_1,i_2,i_3 \leqslant J\\ i_1+i_2+i_3=J}}\frac{{\cal Q}(i_1,i_2,i_3)}{\sqrt{i_1!i_2!}}c_{i_2}c_{i_1} \sum\limits_{l=0}^{M-1}(-1)^l {M-1 \choose l} \sum\limits_{k=0}^{MN+M-1 }\sqrt{k!}\, c_{k}^\dagger  \notag\\&\times [S_{MN+M-1-l-i_3}S_{MN+l-k} +(-1)^{M-1}S_{MN+M-1-l-k}S_{MN+l-i_3}]\ket{\text{Pf}_{N}}\notag\\& -6\sum_{\substack{0  \leqslant  k_1,k_2,i \leqslant J\\ k_1+k_2+i=J}} {\cal Q}(k_1,k_2,i)\frac{c_{i}}{\sqrt{i!}}\sum\limits_{l=0}^{M-1}(-1)^l {M-1 \choose l}S_{MN+M-1-l-k_1}S_{MN+l-k_2}\ket{\text{Pf}_{N}}\notag\\&+ \sum\limits_{l=0}^{M-1}(-1)^l {M-1 \choose l} \sum\limits_{r,k=0}^{MN+M-1 }\sqrt{r!k!}\, c_{r}^\dagger c_{k}^\dagger  Q^{(3\text{bd},\cal Q)}_J S_{MN+M-1-l-r}S_{MN+l-k}\ket{\text{Pf}_{N}}\notag\\
=& 6 Q^{(3\text{bd},\cal Q)}_J\ket{\text{Pf}_{N+2}} -6\sum_{\substack{0  \leqslant  k_1,k_2,i \leqslant J\\ k_1+k_2+i=J}}\frac{{\cal Q}(k_1,k_2,i)}{\sqrt{i!}}c_{i}\sum\limits_{l=0}^{M-1}(-1)^l {M-1 \choose l}S_{MN+M-1-l-k_1}S_{MN+l-k_2}\ket{\text{Pf}_{N}}.
\end{align}\end{widetext}
where we have used \Eq{refor1} in the first step,  $c_{i_3}c_{i_2}c_{i_1}c_{r}^\dagger =\delta_{r,i_1}c_{i_3}c_{i_2}+(-1)^{M-1}\delta_{r,i_2}c_{i_3}c_{i_1}+\delta_{r,i_3}c_{i_2}c_{i_1}+(-1)^{M-1}c_{r}^\dagger c_{i_3}c_{i_2}c_{i_1}$ twice and $c_{r}^\dagger c_{i_2}c_{i_1}=c_{i_2}c_{i_1}c_{r}^\dagger -\delta_{r,i_1}c_{i_2}+(-1)^{M}\delta_{r,i_2}c_{i_1}$ in the second step,  and again \Eq{re1} in the third step in order to condense terms into the first term on the last line. We have also used the identity $Q^{(3\text{bd},\cal Q)}_J S_{MN+M-1-l-r}S_{MN+l-k}\ket{\text{Pf}_{N}}=0$ since $S_{MN+M-1-l-r}$ and $S_{MN+l-k}$ are zero mode generators, and $\ket{\text{Pf}_{N}}$ is assumed to be a zero mode.

Now we need to simplify the last term
\begin{align}\label{last1}&\sum_{\substack{0  \leqslant  k_1,k_2,i \leqslant J\\ k_1+k_2+i=J}}{\cal Q}(k_1,k_2,i)\frac{c_{i}}{\sqrt{i!}} \sum\limits_{l=0}^{M-1}(-1)^l {M-1 \choose l}\notag\\&\times S_{MN+M-1-l-k_1}S_{MN+l-k_2}\ket{\text{Pf}_{N}}.\end{align}
By using the commutator \be[c_{i}, S_{l}]=\sum_{k=1}^{M}(-1)^k {M \choose k} \sqrt{\frac{i!}{(i-k)!}}\,S_{l-k}c_{i-k},\ee this term can be rewritten as \begin{align}&\sum_{m_1,m_2=0}^{M}(-1)^{m_1+m_2}{M \choose m_1}{M \choose m_2}\notag \\&\times\sum_{\substack{0  \leqslant  k_1,k_2,i \leqslant J\\ k_1+k_2+i=J}} \sum\limits_{l=0}^{M-1}(-1)^l    {M-1 \choose l}{\cal Q}(k_1,k_2,i)\notag \\&\times S_{MN+M-1-l-k_1-m_1}S_{MN+l-k_2-m_2}\notag\\&\times\frac{c_{i-m_1-m_2}}{\sqrt{(i-m_1-m_2)!}}\ket{\text{Pf}_{N}}.\end{align}
Under  change of variables, $k_1+m_1+l=k'_1$, $k_2+m_2-l=k'_2$ and $i-m_1-m_2=i'$, the above term will be
\begin{align}&\sum_{m_1,m_2=0}^{M} (-1)^{m_1+m_2} {M \choose m_1}{M \choose m_2}\notag \\&\times\sum\limits_{l=0}^{M-1}(-1)^l {M-1 \choose l}\sum_{\substack{m_1+l  \leqslant  k'_1 \leqslant J+m_1+l\\m_2-l  \leqslant  k'_2 \leqslant J+m_2-l\\-m_1-m_2  \leqslant  i' \leqslant J-m_1-m_2\\ k'_1+k'_2+i'=J}} \notag \\&\times{\cal Q}(k'_1-m_1-l,k'_2-m_2+l,i'+m_1+m_2)\notag \\&\times S_{MN+M-1-k'_1}S_{MN-k'_2}\frac{c_{i'}}{\sqrt{i'!}}\ket{\text{Pf}_{N}}.\end{align}
Similar to \Eq{constraint1}, we shall use a constraint \be \label{scpsiN}
S_\ell c_{i'}\ket{\text{Pf}_{N}}=0  \quad\text{for  $\ell>M(N-1)$.}\ee

As a result of this constraint, the lower limit of  $k'_1$ can be raised to $2M-1$,  the lower limit of  $k'_2$ can be raised to $M$,  the upper limit of   $i'$ can be lowered to $J-(3M-1)$ on account of $k'_1+k'_2+i'=J$.  Also, observe that $i'$ should be nonnegative; therefore, the upper limit of both $k'_1$ and $k'_2$ can be changed to $J$.

After these changes of limits of summations, \Eq{last1} can be finally simplified to
\begin{align}&\sum_{\substack{2M-1  \leqslant  k'_1 \leqslant J\\M\leqslant  k'_2 \leqslant J \\ 0  \leqslant  i' \leqslant J-(3M-1)\\ k'_1+k'_2+i'=J}}\Bigg[\sum_{m_1,m_2=0}^{M} (-1)^{m_1+m_2} {M \choose m_1}{M \choose m_2}\notag \\&\times\sum\limits_{l=0}^{M-1}(-1)^l {M-1 \choose l}\notag\\&\times {\cal Q}(k'_1-m_1-l,k'_2-m_2+l,i'+m_1+m_2)\Bigg]\notag \\&\times S_{MN+M-1-k'_1}S_{MN-k'_2}\frac{c_{i'}}{\sqrt{i'!}}\ket{\text{Pf}_{N}}.\end{align}
As a result of \Eq{combi}, for the summations in the square bracket not to vanish, there should exist at least one term in $\cal Q$ in which the power of $l$, $m_1$  and  $m_2$ should be greater than or equal to $M-1$,   $M$ and  $M$, respectively. However, the degree of $\cal Q$ is less than $3M-1$.  Therefore, the term in the square bracket vanishes, rendering \Eq{last1} zero.

After this lengthy simplification, we obtain $(N+2)Q^{(3\text{bd},\cal Q)}_J\ket{\text{Pf}_{N+2}}=6Q^{(3\text{bd},\cal Q)}_J\ket{\text{Pf}_{N+2}}$.
Therefore, if $\ket{\text{Pf}_{N}}$ is a zero mode of all $Q^{(3\text{bd},\cal Q)}_J$ for some  $N  \geqslant  6 $, so will be $\ket{\text{Pf}_{N+2}}$. By mathematical induction, the fermionic (bosonic) Pfaffian state, as recursively defined in \Eq{refor1}, is thus a zero mode of the three-body Hamiltonian~\eqref{pro}.\hfill $\blacksquare$

\subsection{Root state and filling factor of the fermionic (bosonic) Pfaffian state $\ket{\text{Pf}_N}$}\label{ff}

The Moore-Read FQH state belongs to a large class of trial wave functions that follow a ``root state $+$ squeezing'' paradigm. This holds true for all Jack polynomial FQH  trial states\cite{BH1,BH2, BH2.5, BH3, regnault} and their fermionic counterparts, of which Moore-Read states are examples, and has recently been generalized to a considerable number of mixed Landau level FQH states.\cite{Chen17,Sumanta18,Chen19, Ahari:2022bds, Cruise}
Consider those occupation number eigenstates
$|\{n_i\}\rangle$ in the angular momentum LLL eigenbasis that appear with non-zero coefficient in an $N$-particle state $\ket{\psi}$. $|\{n_i\}\rangle$ is a Slater determinant for fermions and a symmetrized monomial (permanent) for bosons, but we will prefer the neutral term {\em configuration} to refer to both cases. The case of interest will be where $\ket{\psi}$ is a zero mode of the parent Hamiltonian. Then we write

\be\label{rootexpansion}
    \ket{\psi} = \ket{\psi}_{\text{root}} +\sum_{\ket{\{n_i\}}\neq \ket{\psi}_{\text{root}}}
    C_{\{n_i\}} \ket{\{n_i\}}
\ee
where $\ket{\psi}_{\text{root}}$ is comprised of those configurations in the expansion that cannot be obtained from  any other configuration, appearing with nonzero coefficient in $\ket{\psi}$, through so-called inward-squeezing processes\cite{BH2}.
These inward-squeezing processes are generated by the operations
\be\label{inward}
     c^{\dagger}_{j} c^{\dagger}_{i} c_{i-m} c_{j+m}
\ee
where $i \leqslant j$ and  $m > 0$.
Usually, $\ket{\psi}_{\text{root}}$ is proportional to a single configuration such that
all the other configurations in the expansion in \Eq{rootexpansion} can be obtained from it via inward squeezing. However, by our definition, $\ket{\psi}_{\text{root}}$ can also be a linear combination of such configurations, as it may happen that the zero mode $\ket{\psi}$ is a linear combination of simpler zero modes. We refer the reader to the referenced literature\cite{BH1, BH2, BH2.5, BH3,  regnault, ortiz, Chen17,Sumanta18,Chen19, Ahari:2022bds, Cruise} for details. Also, we note that Jack states and their fermonic counterparts have been associated with certain types of recursion relations\cite{regnault,ThomalePRB84:45127,PhysRevB.95.245123}. These are recursion relations for the coefficients of the mode expansion, where particle number is fixed and the recursion proceeds along increasing ``squeezing level'' of the associated modes. This is to be distinguished from the present case, where we defined states recursively in particle number.

 The root states satisfy  Pauli-like principles. In the case of
 a single-component state in a single Landau level, these are known as generalized Pauli principles\cite{BH1, BH2, Chen14}.
 For example, there is no more than one particle in any three consecutive orbitals in root state of $\nu=1/3$ Laughlin state, which corresponds to the familiar $100100100\dotsc$ configuration. The same generalized Pauli principle does, however, apply to other zero modes (not necessarily of the highest density) of the state's parent Hamiltonian.
 For multi-component and/or multi-Landau-level states, our definition of a root state will generally lead to more than one configuration entering $\ket{\psi}_{\text{root}}$, and especially, will lead to root level entanglement. In this case, we speak of  ``entangled'' Pauli principles\cite{Sumanta18}.
 The unprojected $\nu=2/5$ Jain state may serve as an example of this, where this entangled Pauli principle requires next-nearest neighbors to be singlets of an SU(2)-algebra related to the Landau level degrees of freedom, in addition to ruling out double occupancies (with the same angular momentum but different Landau level indices)\cite{Chen17}. Effectively, this leads to a situation where there can be no more than two particles in any five consecutive orbitals, in the root state.
 By contrast,  basis states inward-squeezed from root states do not satisfy these Pauli-like principles.

 As root states contain much information about the universal properties of the underlying state, including statistics\cite{Flavin},
their uses are manifold.
 In an obvious way, they encode the filling fractions of the underlying state, commonly defined as the ratio of the particle number to the highest angular momentum of any orbital occupied in the state (in the thermodynamic limit!).


In this subsection,  we will now prove that
$\ket{\text{Pf}_N}$
has root state \be \label{rootpartition}c_{0}^\dagger  c_{M-1}^\dagger c_{2M}^\dagger  c_{3M-1}^\dagger \dots c_{(N-2)M}^\dagger  c_{(N-1)M-1}^\dagger\ket{0} \ee for even particle number $N$.  This will re-affirm that it has the correct highest occupied orbital (angular momentum $(N-1)M-1$), rendering it the unique densest zero mode of its parent Hamiltonian, thus, identical (up to normalization) to the Moore-Read state at the respective filling factor.
This will also serve to close one loop-hole in the reasoning so far. As for as shown above, it might be possible that the state $\ket{\text{Pf}_N}$ as defined in \Eq{refor1} vanishes, at least for some sufficiently high particle number $N$. We can rule this out below, as we show in particular the state $\ket{\text{Pf}_N}$ has non-zero overlap with the root state~\eqref{rootpartition}.




Again, we prove this by mathematical induction. For $N=2$, the above statement is true, as seen from \Eq{psi2root}.
Now we assume \be\ket{\text{Pf}_{N}}_{\text{root}}\propto c_{0}^\dagger  c_{M-1}^\dagger c_{2M}^\dagger  c_{3M-1}^\dagger \dots c_{(N-2)M}^\dagger  c_{(N-1)M-1}^\dagger\ket{0}\ee for $N\geqslant 2$ and its coefficient $ C_{N_{\text{root}}}$ in the expansion of $ \ket{\text{Pf}_{N}}$  in terms of occupation number basis states is non-zero.

We plug $\ket{\text{Pf}_{N}}_{\text{root}}$ into \Eq{refor} to obtain
\begin{widetext}
\begin{align}
& {\cal R}_N \ket{\text{Pf}_{N}}_{\text{root}} \notag\\
 =& \frac{1}{N+2}\sum\limits_{l=0}^{M-1}(-1)^l {M-1 \choose l} \sum\limits_{r_1,r_2=0}^{MN+M-1}\sqrt{r_1!r_2!}  \,  c_{r_1}^\dagger c_{r_2}^\dagger S_{MN+M-1-l-r_1}S_{MN+l-r_2}\ket{\text{Pf}_{N}}_{\text{root}}\notag\\
=  & \frac{1}{N+2}\sum\limits_{l=0}^{M-1}(-1)^l {M-1 \choose l} \sum\limits_{r_1,r_2=0}^{MN+M-1}\sum\limits_{p_1,\dots p_N,q_1,\dots q_N=0}^{M}(-1)^{\sum_{i=1}^N (p_i+q_i)} \prod_{i=1}^N{M \choose p_i}{M \choose q_i}\notag\\&\times \sqrt{\frac{(q_1+p_1)!}{0!}\frac{(M-1+q_2+p_2)!}{(M-1)!}\frac{(2M+q_3+p_3)!}{(2M)!}\frac{(3M-1+q_4+p_4)!}{(3M-1)!}\dots \frac{[(N-2)M+q_{N-1}+p_{N-1}]!}{[(N-2)M]!}}\notag\\&\times  \sqrt{\frac{[(N-1)M-1+q_N+p_N]!r_1!r_2!}{[(N-1)M-1]!}} \, c_{r_1}^\dagger c_{r_2}^\dagger   c_{q_1+p_1}^\dagger  c_{M-1+q_2+p_2}^\dagger c_{2M+q_3+p_3}^\dagger  c_{3M-1+q_4+p_4}^\dagger \dots  c_{(N-2)M+q_{N-1}+p_{N-1}}^\dagger  \notag\\&\times c_{(N-1)M-1+q_N+p_N}^\dagger   S_{MN+M-1-l-r_1-\sum_{i=1}^N p_i}S_{MN+l-r_2-\sum_{i=1}^N q_i}\ket{0} \notag\\=  & \frac{1}{N+2}\sum\limits_{l=0}^{M-1}(-1)^l {M-1 \choose l} \sum\limits_{p_1,\dots p_N,q_1,\dots q_N=0}^{M}(-1)^{\sum_{i=1}^N (p_i+q_i)} \prod_{i=1}^N{M \choose p_i}{M \choose q_i}\notag\\&\times \sqrt{\frac{(q_1+p_1)!}{0!}\frac{(M-1+q_2+p_2)!}{(M-1)!}\frac{(2M+q_3+p_3)!}{(2M)!}\frac{(3M-1+q_4+p_4)!}{(3M-1)!}\dots\frac{[(N-2)M+q_{N-1}+p_{N-1}]!}{[(N-2)M]!}} \notag\\&\times  \sqrt{\frac{[(N-1)M-1+q_N+p_N]!}{[(N-1)M-1]!}(MN+M-1-l-\sum\nolimits_{i=1}^N p_i)!(MN+l-\sum\nolimits_{i=1}^N q_i)!}\notag\\&\times  c_{MN+M-1-l-\sum_{i=1}^N p_i}^\dagger c_{MN+l-\sum_{i=1}^N q_i}^\dagger    c_{q_1+p_1}^\dagger  c_{M-1+q_2+p_2}^\dagger c_{2M+q_3+p_3}^\dagger  c_{3M-1+q_4+p_4}^\dagger \dots c_{(N-2)M+q_{N-1}+p_{N-1}}^\dagger \notag\\&\times   c_{(N-1)M-1+q_N+p_N}^\dagger  \ket{0},\notag
\end{align}\end{widetext}
where we have used \Eq{scdag} to move $S$ to the right of $c^\dagger$.
We  have also used the fact that both indices of $S$ operators,  $MN+M-1-l-r_1-\sum_{i=1}^N p_i$
and $MN+l-r_2-\sum_{i=1}^N q_i$  have to be 0, following the same logic used in the derivation of \Eq{psi2root}.

The only solutions for
\begin{align}& c_{MN+M-1-l-\sum_{i=1}^N p_i}^\dagger c_{MN+l-\sum_{i=1}^N q_i}^\dagger   c_{q_1+p_1}^\dagger  c_{M-1+q_2+p_2}^\dagger \notag
\\&\times c_{2M+q_3+p_3}^\dagger  c_{3M-1+q_4+p_4}^\dagger \dots c_{(N-2)M+q_{N-1}+p_{N-1}}^\dagger   \notag
\\&\times   c_{(N-1)M-1+q_N+p_N}^\dagger  \ket{0}
\end{align} in the above expression to be proportional to $\ket{\text{Pf}_{N+2}}_{\text{root}}\propto c_{0}^\dagger  c_{M-1}^\dagger c_{2M}^\dagger  c_{3M-1}^\dagger \dots c_{NM}^\dagger  c_{(N+1)M-1}^\dagger\ket{0}$ are parameterized by a choice of $j=0,1,2,\dots,N/2$, where
$ q_1=p_1=q_2=p_2=\dots= q_{2j}= p_{2j}=0$, and $ q_{2j+1}=p_{2j+1}=q_{2j+2}=p_{2j+2}=\dots= q_{N}= p_{N}=M$,   and furthermore a choice of $l=0,M-1$.
One checks that all these solutions enter with the same sign,
and thus,
$\ket{\text{Pf}_{N+2}}_{\text{root}}$ will be generated from $\ket{\text{Pf}_{N}}_{\text{root}}$ via \Eq{refor}.
On the other hand,
by acting with  ${\cal R}_N$
on any $\ket{\{n_i\}}$
that can be obtained from
$ \ket{\text{Pf}_{N}}_{\text{root}}$
via inward squeezing, similar considerations show that $\ket{\text{Pf}_{N+2}}_{\text{root}}$ cannot be generated, and the only configurations that can be generated are obtainable from $ \ket{\text{Pf}_{N+2}}_{\text{root}}$
via inward squeezing.
Together, these results show that
$ \ket{\text{Pf}_{N+2}}_{\text{root}}$
is  the root state
of $\ket{\text{Pf}_{N+2}}$ not only in name, but according to the definition given at the beginning of this section.


In summary, the fermionic (bosonic) Pfaffian state $\ket{\text{Pf}_N}$, as recursively defined in \Eq{refor} for even particle number $N$,  has a 
root state proportional to \be\label{evenNPfroot} c_{0}^\dagger  c_{M-1}^\dagger c_{2M}^\dagger  c_{3M-1}^\dagger \dots c_{(N-2)M}^\dagger  c_{(N-1)M-1}^\dagger\ket{0}\,, \ee thus possessing the filling factor $1/M$.

\subsection{Off-diagonal long-range order operator of Pfaffian state in second quantization}\label{fracc}
In this subsection, we establish the connection between the foregoing results and
existence of off-diagonal long-range order (ODLRO)
in a non-local order parameter for the Moore-Read state.
Such a connection is natural, as the second-quantized recursion  \eqref{refor} we use to define the Moore-Read state in this paper is a generalization of a similar recursion for the Laughlin state that, in its original form\cite{ReadOP}, emerged as the interpretation of the Laughlin state as a condensate of a non-local order parameter. This is quite manifest also in \Eq{refor}, and can be further  emphasized by its trivial formal ``integration'' via
\begin{equation}\label{refor3}
\ket{\text{Pf}_{N}}=({\cal R})^{N/2}\ket{0}
\end{equation}
for $N$ even, where
\begin{equation}
{\cal R} = \sum_{N \text{even}} {\cal R}_N P_N\,,
\end{equation}
and $P_N$ is the projection onto $N$-particle subspace of the Fock space.
In this form,
one may see this equation to be equivalent to Eq. (5.8) by Moore and Read\cite{MR}, with the important difference that the latter is presented in mixed first/second-quantized notations.

Fully second-quantized forms similar to ours have been given before for the Laughlin state\cite{Chen14},
concurrent with second-quantized expressions for the associated non-local order parameter\cite{Mazaheri14}. Both have been successfully generalized to
 composite fermion states\cite{Chen19}, which became instrumental in constructing parent Hamiltonians for these states\cite{PhysRevLett.124.196803}.
 To complete our second-quantized picture for the Moore-Read state, it is thus prudent to construct the non-local order parameter directly and demonstrate its display of ODLRO.
 Similar to previously studied examples, the key ingredient is the action of an electron destruction operator on the incompressible ground state, as facilitated in the present case by \Eq{re1}. While
 Refs. \onlinecite{Mazaheri14, Chen19}
demonstrated the ODLRO in the orbital basis, a formulation in real space is equally possible. We will aim for the demonstration of real-space ODLRO here, and to this end, utilize some notation developed in Ref. \onlinecite{2021exact}.


We thus introduce
the field operator annihilating a particle (we again treat fermion and boson on equal footing) at $z =x+i\,y$, projected onto the lowest Landau level, via its mode expansion
 $\Lambda(z)=\sum_{r\geqslant 0} \phi_r(z) c_r$, where the single-particle wave function on the disk is \be\label{spwfdisk}\phi_r(z) =N_r^{-1}z^re^{-|z|^2/4}\ee  with the normalization factor $N_r=\sqrt{2\pi 2^r r!\,}$. By introducing pseudo-fermionic (bosonic) operators\cite{Chen19} $\bar{c}_r:=c_r/N_r$ and $\bar{c}_r^\dagger:=N_r c_r^\dagger$ for compactness, \Eq{re1} can be recast in the form \begin{align}
\Lambda(z) \ket{\text{Pf}_{N+2}}=&\frac{e^{-|z|^2/4}}{4\pi}\sum\limits_{l=0}^{M-1}(-1)^l {M-1 \choose l}\sum \limits_{r,k\geqslant 0} \frac{z^r \bar{c}_{k}^\dagger}{\sqrt{2^{r+k}}}  \notag \\ &\times [S_{MN+M-1-l-r}S_{MN+l-k}  +(-1)^{M-1} \notag\\ & \times S_{MN+M-1-l-k}S_{MN+l-r}]\ket{\text{Pf}_{N}}.\end{align}

This may be simplified by introducing the second-quantized $N$-body quasihole operator $\widehat{U}_N(z)=\sum_{d=0}^N (-z)^{N-d}2^{\frac{d}{2}}e_d$, which creates a Laughlin-quasihole at $z$. Its $M$th power is given by $\widehat{U}^M_N(z)=(-1)^{MN}\sum_{r\geqslant 0}z^r 2^{\frac{MN-r}{2}} S_{MN-r}$ (see supplementary notes of Ref. \onlinecite{2021exact}).
Note that Read's order parameter for the $1/M$-Laughlin state is precisely
$\Lambda^\dagger(z) \widehat{U}^M_N(z)$, albeit with the role of fermions and bosons reversed compared to the present case.
Using the commutativity of the $S$-operators among themselves, and the fact that \be S_m \ket{\text{Pf}_{N}}=0  \quad\text{for  $m>MN$,}\ee we now obtain
\begin{align}\Lambda(z) \ket{\text{Pf}_{N+2}}=\mathcal{F}_{M,N}(z)\ket{\text{Pf}_{N}},\end{align} where
\begin{align}\mathcal{F}_{M,N}(z)=&\frac{(-1)^{MN}  e^{-|z|^2/4} }{2\pi\sqrt{2^{MN+M-1}}}\sum\limits_{l=0}^{M-1}(-1)^l {M-1 \choose l}\notag\\&\times z^{M-1-l}\sum\limits_{k \geqslant 0}   \frac{\bar{c}_{k+l}^\dagger}{\sqrt{2^k}}  S_{MN-k}\widehat{U}^M_N(z).\end{align}



In line with Read's original reasoning for the Laughlin state,\cite{ReadOP} we can argue that
\begin{align}\label{eqODLRO}&\bra{\text{Pf}_{N}}\mathcal{F}^\dagger_{M,N} (z) \Lambda (z)\Lambda^\dagger(z') \mathcal{F}_{M,N} (z') \ket{\text{Pf}_{N}}\notag\\=&\bra{\text{Pf}_{N+2}}\rho (z) \rho (z') \ket{\text{Pf}_{N+2}}\notag\\ \rightarrow&\langle\rho\rangle ^2,\end{align} where we use the Landau-level projected fields $\Lambda(z)$
to define local densities $\rho(z)=\Lambda^\dagger(z)\Lambda(z)$, such that $\hat N =\int d^2z \,\rho(z)$ is the Landau-level projected particle number operator. We also
assumed the exponential decay of correlations as $\lvert z-z'\rvert\rightarrow\infty$,
such that the expression
approaches the square of the particle density $\langle\rho\rangle$ of the homogeneous fluid, which is determined by the filling factor $\nu$.

We thus infer the existence of ODLRO of the $\nu=1/M$ Moore-Read Pfaffian state
in the non-local operator given by
\be\mathcal{O} (z)=\Lambda^\dagger(z) \mathcal{F}_{M,N} (z).\label{order}\ee
It is worth noting that,
in spite of deliberately writing \eqref{order} in a form similar to the Laughlin-state order parameter $\Lambda^\dagger(z) \widehat{U}^M_N(z)$, there are important differences. The most crucial difference lies in the fact that
\Eq{order} changes particle number by $2$, as a change by $1$ is also ``hidden'' in the field operator $\mathcal{F}_{M,N} (z)$. The fact that the order parameter changes the particle number by $2$ is, of course, a direct signature of the paired nature of the Moore-Read state.
We emphasize once more that the presentation of the Moore-Read state in the form \eqref{refor3} is by itself not sufficient to demonstrate ODLRO. For this, we crucially needed \Eq{re1}.

Given the above, following again Read's construction\cite{ReadOP}, we could alternatively use \Eq{refor3}
(together with \Eq{re1})
to construct a condensate of a well-defined phase conjugate to particle number, for which the order parameter \eqref{order} itself assumes an expectation value. The only difference with the Laughlin-state case would be that such a condensate would have well-defined particle number parity, i.e., it would be a coherent superposition of states \eqref{refor3} with even $N$ only. We leave the (simple) details to the reader.

\subsection{Higher angular momentum paired Pfaffian states}\label{HigherPfaffian}
We generalize the results of Section \ref{MR2ndquant} to Pfaffian state of the form
\be\label{mwavemain}\Psi_{N}^m \sim \text{Pf}\left[\frac{1}{(z_i-z_j)^m}\right]\prod\limits_{1\leqslant i<j\leqslant N}(z_i-z_j)^M,\ee
where odd $m\leqslant M$. $\text{Pf}\left[\frac{1}{(z_i-z_j)^m}\right]$ signifies paired composite fermions beyond $p$-wave pairing, where in particular the case $m=3$ has recently been studied\cite{fwave}.

For this state, the recursion relation Eq. \eqref{refor} generalizes straightforwardly via the modification ${\cal R_N}\rightarrow {\cal R}^m_N$, where
\begin{subequations}\label{generalPfm}
\begin{align}\label{RmN}
{\cal R}^m_N = \frac{1}{N+2}\sum\limits_{l=0}^{M-m}& (-1)^l {M-m \choose l} \sum\limits_{r,k=0}^{MN+M-m }\sqrt{r!k!}\, c_{r}^\dagger c_{k}^\dagger \notag\\&\times  S_{MN+M-m-l-r}S_{MN+l-k}\,,
\end{align}
such that
\begin{align}\label{refor2m}
\ket{\text{Pf}_{N+2}^m}={\cal R}^m_N\ket{\text{Pf}_{N}^m}\,,
\end{align}
\end{subequations}
where we also introduced a ket
$\ket{\text{Pf}_{N}^m}$ associated with the wave function \eqref{mwavemain}.

For the state \eqref{mwavemain}, we do no know an appropriate parent Hamiltonian at this point, so the proof of Eq. \eqref{generalPfm} necessarily proceeds by making contact with the first-quantized form given in Eq. \eqref{mwavemain}. This is done in Appendix \ref{1stproof}, where we also specify pertinent normalization conventions.
Equally importantly, one can generalize the effect of particle removal, \Eq{re1}, as follows
\begin{align}\label{re1m}
c_r \ket{\text{Pf}^m_{N+2}}=&\frac{\sqrt{r!}}{2}\sum\limits_{l=0}^{M-m}(-1)^l {M-m \choose l} \sum\limits_{k=0}^{MN+M-m }\sqrt{k!}\, c_{k}^\dagger\notag\\&\times   [S_{MN+M-m-l-r}S_{MN+l-k}+(-1)^{M-m}\notag \\ & \times  S_{MN+M-m-l-k}S_{MN+l-r}]\ket{\text{Pf}^m_{N}}.\end{align}
A derivation of \Eq{re1m} from the first-quantized \Eq{mwavemain} is again given in Appendix \ref{1stproof}, or, from the second-quantized \Eq{generalPfm}, in the Supplemental Material\cite{Supplement}. The benefit of \Eq{re1m} is, among other things, a straightforward generalization of the derivation of ODLRO given in the preceding section to the case of \Eq{mwavemain}.
This leads to ODLRO in the following non-local operator,
\be\mathcal{O} (z)=\Lambda^\dagger(z) \mathcal{F}_{M,m,N} (z),\ee
where
\begin{align}\mathcal{F}_{M,m,N}(z)=&\frac{(-1)^{MN}  e^{-|z|^2/4} }{2\pi\sqrt{2^{MN+M-m}}}\sum\limits_{l=0}^{M-m}(-1)^l {M-m \choose l}\notag\\&\times z^{M-m-l}\sum\limits_{k \geqslant 0}   \frac{\bar{c}_{k+l}^\dagger}{\sqrt{2^k}}  S_{MN-k}\widehat{U}^M_N(z).\end{align}
We leave other possible uses of \Eq{re1m}, such as in the construction of possible parent Hamiltonians for \Eq{mwavemain}, to future work.

\section{Recursive formula for fermionic $\nu=1/2$ anti- and PH-Pfaffian states}\label{aP}

At Landau level filling factor
$\nu=1/2$, several inequivalent topological phases featuring Majorana fermions are possible. Among possible competitors, the
anti-Pfaffian state has been proposed as the particle-hole conjugate of $\nu=1/2$ Pfaffian state\cite{levinAP,leeAP}.  Generally, a particle-hole conjugate of a state can be obtained by replacement $c\rightarrow h^\dagger$, $c^\dagger \rightarrow h$, and $\ket{0}_e \rightarrow \prod_{i=0}^{l_{\sf max}(N)} h_i^\dagger\ket{0}_h$, where $l_{\sf max}(N)$ is the highest occupied orbital in the $\nu=1/2$ Pfaffian state, $l_{\sf max}(N)= 2N-3$ for $N$ even. As long as we restrict ourselves to the Fock space associated with the orbitals $0,\dotsc, l_{\sf max}(N)$, these relations merely facilitate a re-interpretation of the Pfaffian state. A new state is obtained when the ``holes'' created by the operators $h^\dagger$ are again re-interpreted as the particles (i.e., once more replaced by $c^\dagger$'s). We leave this understood. On the half-infinite lattice, however, the replacement $\ket{0}_e \rightarrow \prod_{i=0}^{l_{\sf max}(N)} h_i^\dagger\ket{0}_h$ does change the vacuum. It replaces the ``particle vacuum'' for orbitals with angular momenta $l>l_{\sf max}(N)$ with the ``hole vacuum'', i.e., a $\nu=1$ integer quantum Hall state. The result is that once the $h^\dagger$-operators
are re-interpreted as particles, we obtain the $(N-2)$-particle anti-Pfaffian state $\ket{\text{aPf}_{N-2}}$ from the $N$-particle $\nu=1/2$ Pfaffian state  $\ket{\text{Pf}_{N}}$, where  $\ket{\text{aPf}_{N-2}}$ has the same highest occupied orbital $l_{\sf max}(N)= 2N-3$, and has an edge with vacuum. The following example illustrates this:
The four-particle Pfaffian state on the disk is
$(c_{0}^\dagger  c_{1}^\dagger c_{4}^\dagger  c_{5}^\dagger -\sqrt{2}c_{0}^\dagger  c_{2}^\dagger c_{3}^\dagger  c_{5}^\dagger+\sqrt{10}c_{1}^\dagger  c_{2}^\dagger c_{3}^\dagger  c_{4}^\dagger)\ket{0}_e$. After replacement $c\rightarrow h^\dagger$, $c^\dagger \rightarrow h$, and $\ket{0}_e \rightarrow \prod_{i=0}^5 h_i^\dagger\ket{0}_h$, we obtain two-particle anti-Pfaffian state on the disk $(h_{2}^\dagger  h_{3}^\dagger  -\sqrt{2}\, h_{1}^\dagger  h_{4}^\dagger+\sqrt{10}\, h_{0}^\dagger  h_{5}^\dagger)\ket{0}_h$.
We note that $l_{\sf max}(N)$ agrees with the number of flux quanta on the sphere the respective state would require to represent a rotationally invariant state.

Using the above, by particle-hole conjugating the second-quantized recursive formula \Eq{refor1} from $(N+2)$-particle fermionic $\nu=1/2$ Pfaffian state to $(N+4)$-particle state with $M=2$, we can arrive at the second-quantized recursive formula for the fermionic $\nu=1/2$ anti-Pfaffian (aPf) state,
\begin{align}\label{anrec}
\ket{\text{aPf}_{N+2}}=& \frac{2}{N+4}\sum\limits_{r,k=0}^{2N+5}\sqrt{r!k!}\,   h_{r}  h_{k} R_{2N+5-r}R_{2N+4-k}\notag\\&\times h_{2N+2}^\dagger h_{2N+3}^\dagger h_{2N+4}^\dagger h_{2N+5}^\dagger\ket{\text{aPf}_{N}}\,,\end{align}  for even nonnegative $N$.
The beginning of recursion is $\ket{\text{aPf}_{0}}=\ket{0}_h$, the vacuum for holes. Four hole creation operators appear in the recursive formula, since each time we increase the particle number by two, the ``edge'' between vacuum and $\nu=1$ phase in the vacuum replacement $\ket{0}_e \rightarrow \prod_{i=0}^{l_{\sf max}(N)} h_i^\dagger\ket{0}_h$ shifts by four orbitals.
The $R$ operator in the above recursive formula for the anti-Pfaffian state is obtained from $S$ operator in \Eq{S} with $M=2$ by particle-hole conjugation.
Explicitly, \be\label{Rop} \begin{split}
& {R_\ell } = {( - 1)^\ell }\sum\limits_{n_1 + n_2= \ell }  f_{n_1} f_{n_2}\quad \text{for} \quad \ell\geqslant 0,\\&
R_\ell=0\quad \text{for} \quad \ell<0.\end{split}\ee
Here,   $f_n$ is the particle-hole conjugate of $e_n$ in \Eq{e},
\begin{align}\label{fn}
f_n =& \frac{1}{n!} \sum_{l_1,\dots, l_n=  0}^{ + \infty } \sqrt {l_1 + 1} \, h_{l_1+1} \sqrt {l_2+ 1} \, h_{l_2 + 1}  \cdots\notag\\&\times \sqrt {l_n+ 1}h_{ l_n+ 1}    h^\dag_{l_n} \cdots h^\dag_{l_2}h^\dag_{l_1}\quad \text{for}~n>0,\notag\\
f_0=&\mathbb{1},\notag\\ f_n=&0 \quad \text{for}~n<0.\end{align} Note that different $R_\ell$ still commute with each other. For $\ell >0$, $S_\ell$ increases the total angular momentum of an electronic state by $\ell$, whereas its particle-hole conjugate $R_\ell$ decreases the total angular momentum, as measured by occupied $h^\dagger$-states, by the same amount.

The parent Hamiltonian for $N$-particle anti-Pfaffian state is the particle-hole conjugate of the three-body parent Hamiltonian for $\nu=1/2$ Pfaffian state in \Eq{pro} with $M=2$,

\be H_{\text{aPf}_N}\label{PHaaPf}=\sum\limits_{J} U^{\dag}_{J,N}  U_{J,N}\ee with
\begin{align} U_{J,N}= \sum_{\substack {i_1+i_2+i_3=J\in [3,6N]}}&\frac{\sqrt{6(J-3)!}}{3^{\frac{J}{2}}4\sqrt{i_1!i_2!i_3!}}(i_1-i_2)(i_1-i_3)\notag\\&\times (i_2-i_3) h^\dagger_{i_3}h^\dagger_{i_2}h^\dagger_{i_1}.
\end{align}
Note that, however, the above Hamiltonian has $N$-particle anti-Pfaffian state as the unique incompressible zero mode only if orbital indices in the above sum are restricted by the additional constraint ${0  \leqslant  i_1,i_2,i_3 \leqslant  2N+1}$,
{\em or} if the edge with ($h$-)vacuum is instead replaced with an edge with a $\nu=1$ state. This is the reason why the edge of the anti-Pfaffian with vacuum is more complicated than that of the original Moore-Read state.\cite{levinAP,leeAP}

We remark that although the case $M=2$ is of greatest interest, one may generalize the above straightforwardly to obtain recursions
for the particle-hole conjugates of $\nu=1/M$ Moore-Read states,
although these would then not live at the same filling factor in the thermodynamic limit, but instead would have filling factor $1-1/M$.

Note moreover that by straightforwardly taking the particle-hole conjugate of \Eq{order}, we may define non-local order parameters for these particle-hole conjugates of Moore-Read states, as arguments leading to  \Eq{eqODLRO} will, mutatis mutandis, hold. In particular, by particle-hole conjugation of \Eq{re1}, one obtains a similar equation for particle addition into the particle-hole conjugate of Moore-Read states.

While so far, we have mostly focused on states at even particle number $N$, we can easily obtain the incompressible Moore-Read state at odd particle number $N$ via
\be\label{Pfodd}
\ket{\text{Pf}_N} = c_{l_{\sf max}(N+1)} \ket{\text{Pf}_{N+1}}\,.
\ee
Note that for general $M$, $l_{\sf max}(N) = M(N-1)-1$ for $N$ even,  $l_{\sf max}(N) = M(N-1)$ for $N$ odd. (See \Eq{evenNPfroot})
For odd $N$, the particle-hole conjugate of  $\ket{\text{Pf}_N}$ has $l_{\sf max}(N)+1-N = MN-M-N+1$ particles within the orbitals $0,1,2,\dotsc, l_{\sf max}(N)$, which is also even. (Note that we are dealing with fermionic states in this section, so $M$ is even)
It is thus more natural to define the $\nu=1/2$
anti-Pfaffian ($M=2$) for odd $N$ in analogy with \Eq{Pfodd} via
\be
  \ket{\text{aPf}_N}= c_{l_{\sf max}(N+3)} \ket{\text{aPf}_{N+1}}\,,
\ee
since $(N+1)$-particle $\nu=1/2$ anti-Pfaffian state is obtained  from the $(N+3)$-particle $\nu=1/2$ Pfaffian state  by particle-hole conjugation.

Lastly, the PH-Pfaffian phase
recently attracted much interest,\cite{Son,Zucker,Jolicoeur2007}
which is the universality class of a
particle-hole symmetric state at $\nu=1/2$.  Inspired by the latter and with the help of the above developments, we may easily construct a \textit{particle-hole symmetric} state
defined by straightforward
modification and amalgamation of the recursions  for the $\nu=1/2$ Pfaffian and anti-Pfaffian states,
\begin{align}\label{PHrec}
\ket{\text{PH}_{N+2}}=&  \sum\limits_{r,k=0}^{2N+3}\sqrt{r!k!}\,( c_{r}^\dagger c_{k}^\dagger S_{2N+3-r}S_{2N-k}+c_{r}  c_{k} \notag\\&\times R_{2N+3-r}R_{2N-k}c_{2N}^\dagger c_{2N+1}^\dagger c_{2N+2}^\dagger c_{2N+3}^\dagger  )\notag\\&\times\ket{\text{PH}_{N}}\,,\end{align}  for even nonnegative $N$.
The beginning of recursion is given by $\ket{\text{PH}_{0}}=\ket{0}$, the vacuum for electrons. The state
$\ket{\text{PH}_N}$ so constructed is manifestly particle-hole symmetric
on the orbital lattices given by the orbitals with indices $0,\dotsc, 2N+3$. In particular, $\ket{\text{PH}_{N+2}}$ would thus suitably fit onto a sphere
with the correct number of flux quanta $2(N+2)-1$.
In the above, the $R$ operator is still defined as in \Eq{Rop}, but with all $h$-operators in $f_n$ replaced by $c$-operators, as they must be creating the same particles as those in the $S$-operator part of the recursion.

We defer further  analysis of the state defined in \Eq{PHrec} and its relation to the first-quantized particle-hole symmetric Pfaffian state defined in the literature\cite{Zucker,PhysRevB.98.081107,PhysRevB.102.195153,PhysRevB.104.245306}, or possibly a gapless particle-hole symmetric state at half-filling\cite{Son}, to future work.

\section{Discussion and outlook}\label{DO}

In this paper, we developed a second-quantized presentation for the Moore-Read state at filling factor $\nu=1/M$. In practice, this presentation is realized as a recursive definition of Moore-Read states in second quantization. Such recursions are of interest in connection with the recent body of literature  about the construction of frustration-free parent Hamiltonians for FQH states in second quantization, which can, in principle, lead to new Hamiltonians that are difficult to construct following the established first-quantized principles. The prime example for such a development is given by the recently
constructed Hamiltonians for the (positive) Jain sequence\cite{PhysRevLett.124.196803}. Two types of presentations for fractional quantum Hall trial wave functions can be distinguished that are both far removed from traditional first-quantized constructions and lend themselves to the scheme for the discussion of parent Hamiltonians
that is the subject of this
paper. One is the MPS-presentation of fractional quantum Hall trial wave functions, which also exists for Moore-Read states, but not, to our knowledge, for composite fermion states or the anti-Pfaffian state. The other consists in recursion relations that are closely related to an understanding of the state in question as a condensate of a non-local order parameter. The latter kind of presentation is what we utilized and further developed in this work for the Moore-Read states. A closely related mixed first/second-quantized presentation of this kind has been known for some time\cite{MR}. While we give a fully second-quantized version of this presentation, this, by itself, was not sufficient for the second-quantized discussion of parent Hamiltonians we have given in this work. Instead, a key ingredient developed in this paper is  the second-quantized description of particle removal from this state in the form of \Eq{re1}. On the one hand, this allows us to develop a fully second-quantized understanding of the parent Hamiltonian of Moore-Read states. As the example of the composite fermion states shows, such an understanding furnishes a promising foundation on which to base the construction of new parent Hamiltonians that are not based on simple clustering properties manifest in first quantization.
Moreover, \Eq{re1}  also makes possible our derivation of off-diagonal long-range order in these states, in terms of non-local order parameters. We have also shown how both the second-quantized presentation as well as the definition of the  non-local order parameter extend to particle-hole conjugates of Moore-Read states.
Some of our findings are complementary to similar developments utilizing MPS presentation of Moore-Read states\cite{SchosslerChenSeidel_tobepublished}.  We are hopeful that these findings will continue to facilitate developments of trial fractional quantum Hall states and accompanying parent Hamiltonians that are not conveniently available in the traditional first-quantized approach. Moreover, the distinction
between various similar non-Abelian phases at half-filling has
inspired several proposals in the past, guiding both physical\cite{HuberPRL94:016805, SeidelPRB80:241309, WangB81:035318} and numerical experiment\cite{Zucker,PhysRevLett.119.026801,Parton_antiPfaffian,PhysRevB.98.081107,PhysRevB.101.041302,PhysRevLett.125.146802,PhysRevB.102.195153,PhysRevB.104.L081407,PhysRevB.104.245306}.
We hope that the formulas we developed here for non-local order parameters can provide additional tools to distinguish the underlying states at least in numerical experiments.

Note added: While preparing this manuscript, we became aware of a work in parallel by A. Bochniak and G. Ortiz,\cite{Bochniak_tobepublished} which contains a second-quantized presentation of the Moore-Read states equivalent to ours, but otherwise focuses on different aspects of the physics of these states.



\begin{acknowledgments}
L.C. is supported by NSFC Grant No. 12004105. A. S.  acknowledges support by the National Science Foundation under Grant No. DMR-2029401.
We gratefully acknowledge insightful discussions with G. Ortiz, A. Bochniak, and A. Balram. 
\end{acknowledgments}
\appendix
\begin{widetext}

\section{The derivation of Eqs.~\eqref{refor1} and~\eqref{re1} in first quantization}\label{1stproof}
We can write Moore-Read's (unnormalized) first-quantized Pfaffian   wave function as
\be\Psi_{N}={\cal N}_N\,\text{Pf}\left(\frac{1}{z_i-z_j}\right)\prod\limits_{1\leqslant i<j\leqslant N}(z_i-z_j)^M,\ee
with  even (odd) $M$ for fermions (bosons) and even particle number $N$, and an as yet arbitrary normalization constant ${\cal N}_N$. We will fix the normalization convention below.

Moore-Read's original Pfaffian state   has also been generalized to an $f$-wave paired state of first-quantized wave function\cite{fwave}
\be\text{Pf}\left[\frac{1}{(z_i-z_j)^3}\right]\prod\limits_{1\leqslant i<j\leqslant N}(z_i-z_j)^M,\ee which inspires us to consider a generalized  Pfaffian state
\be\label{mwave}\Psi_{N}^m={\cal N}_N^m\,\text{Pf}\left[\frac{1}{(z_i-z_j)^m}\right]\prod\limits_{1\leqslant i<j\leqslant N}(z_i-z_j)^M,\ee with an odd positive integer $m$ as the pairing parameter, which must obey $1\leqslant m \leqslant M$
and on which the normalization
${\cal N}_N^m$ may depend.

In all of the above, Pf is the Pfaffian of an antisymmetric matrix with element $1/(z_i-z_j)^m$,
\be\text{Pf}\left[\frac{1}{(z_i-z_j)^m}\right]=\frac{1}{2^{\frac{N}{2}}(\frac{N}{2})!}\sum\limits_{\sigma\in S_N}(-1)^{\sigma} \prod\limits_{k=1}^{\frac{N}{2}}\frac{1}{(z_{\sigma_{2k-1}}-z_{\sigma_{2k})^m}}.\ee
The permutation $\sigma$ can be viewed as encoding a way of pairing indices into pairs  $(\sigma_{2k-1},\sigma_{2k})$. There is then, however, much overcounting, as both the order within pairs and between pairs does not matter. This is compensated by a factor $\frac{1}{2^{N/2}(N/2)!}$. As the order of pairs plays no role, we can, in particular, still generate all pairings if we fix $\sigma_{N}=N$. We write such $\sigma$ as $\sigma\in S_{N-1}$. Thus, adjusting the combinatorial overcounting factor,
\begin{align}\text{Pf}\left[\frac{1}{(z_i-z_j)^m}\right]=&\frac{1}{2^{\frac{N-2}{2}}(\frac{N-2}{2})!}\sum\limits_{\sigma\in S_{N-1}}(-1)^{\sigma}\prod\limits_{k=1}^{\frac{N}{2}}\frac{1}{(z_{\sigma_{2k-1}}-z_{\sigma_{2k}})^m}\notag\\
=&\frac{(N-1)!}{2^{\frac{N-2}{2}}(\frac{N-2}{2})!}\mathcal{A}_{N-1}\prod\limits_{k=1}^{\frac{N}{2}}\frac{1}{(z_{2k-1}-z_{2k})^m},\end{align}
where $\mathcal{A}_{N-1}$ denotes the antisymmetrization  in just $z_1,\cdots,z_{N-1}$. Thus,
\be\Psi_{N}^m={{\cal N}^m_N}' \,\prod\limits_{1\leqslant i<j\leqslant N}(z_i-z_j)^M\mathcal{A}_{N-1}\prod\limits_{k=1}^{\frac{N}{2}}\frac{1}{(z_{2k-1}-z_{2k})^m},\ee
where we have absorbed all combinatorial factors into a new normalization constant
${{\cal N}^m_N}'$.

For even $M$, the Laughlin-Jastrow factor is totally symmetric, we can pull it into the antisymmetrization. For  odd $M$, the Laughlin-Jastrow factor is totally antisymmetric, and we
can change the anti-symmetrization into a symmetrization after pulling the Laughlin-Jastrow factor inside.
 We thus define $\mathcal{S}^{(M)}_N$ to be the (anti)symmetrization operator in  $z_1,\cdots,z_N$ for  (even) odd  $M$. Changing from $N$ to $N+2$:
\begin{align}
\Psi_{N+2}^m &={{\cal N}_{N+2}^m}' \,\mathcal{S}^{(M)}_{N+1}\,\prod\limits_{1\leqslant i<j\leqslant {N+2}}(z_i-z_j)^M \prod\limits_{k=1}^{\frac{N+2}{2}}\frac{1}{(z_{2k-1}-z_{2k})^m}\notag\\
&={{\cal N}_{N+2}^m}' \,\mathcal{S}^{(M)}_{N+1}(z_{N+1}-z_{N+2})^{M-m}\prod\limits_{1\leqslant i\leqslant N}(z_{N+2}-z_i)^M\prod\limits_{1\leqslant i\leqslant N}(z_{N+1}-z_i)^M \notag\\
&\times\mathcal{S}^{(M)}_{N-1}\prod\limits_{1\leqslant i< j\leqslant N}(z_i- z_j)^M\prod\limits_{k=1}^{\frac{N}{2}}\frac{1}{(z_{2k-1}-z_{2k})^m}.\end{align}

In the above, it does not hurt to insert an additional (anti)symmetrization operator $\mathcal{S}^{(M)}_{N-1}$ in front of the last line as shown, because the
products in the first line are already symmetric in the variables $z_i$ for $i=1\dotsc N$, whereas the second line depends on only these variables;
we were thus able to write
$\mathcal{S}^{(M)}_{N+1}=\mathcal{S}^{(M)}_{N+1} \mathcal{S}^{(M)}_{N-1}$, and permute the $\mathcal{S}^{(M)}_{N-1}$ to the position shown.
This gives
\be\Psi_{N+2}^m=
\frac{{{\cal N}_{N+2}^m}'}{{{\cal N}_{N}^m}'}\,
\mathcal{S}^{(M)}_{N+1}(z_{N+1}-z_{N+2})^{M-m}\prod\limits_{1\leqslant i\leqslant N}(z_{N+2}-z_i)^M\prod\limits_{1\leqslant i\leqslant N}(z_{N+1}-z_i)^M \Psi_{N}^m,\ee
Now we need to expand $\prod\limits_{1\leqslant i\leqslant N}(z_{N+2}-z_i)^M$. To do so, we first expand
\begin{align}\prod\limits_{1\leqslant i\leqslant N}(z_{N+2}-z_i)=&\sum\limits_{k=0}^N  z_{N+2}^k (-1)^{N-k}\sum\limits_{1\leqslant i_1<i_2\cdots<i_{N-k}\leqslant N} z_{i_1}z_{i_2}\cdots z_{i_{N-k}}\notag\\=&\sum\limits_{k=0}^N  z_{N+2}^k (-1)^{N-k}\,2^{\frac{N-k}{2}}e_{N-k},\end{align} where we have identified $\sum\limits_{1\leqslant i_1<i_2\cdots<i_{N-k}\leqslant N} z_{i_1}z_{i_2}\cdots z_{i_{N-k}}$ as  $2^{\frac{N-k}{2}}e_{N-k}$.
Then we have \be\prod\limits_{1\leqslant i\leqslant N}(z_{N+2}-z_i)^M=\sum\limits_{k=0}^{MN}  z_{N+2}^k \,2^{\frac{MN-k}{2}}S_{MN-k}, \ee where $S$ is related to $e$ by \Eq{S}. $\prod\limits_{1\leqslant i\leqslant N}(z_{N+1}-z_i)^M$ is expanded in the same way.  $(z_{N+1}-z_{N+2})^{M-m}$ can be expanded via binomial expansion.

With these expansions, we obtain
\be\label{nplus2}\Psi_{N+2}^m=\frac{1}{2\pi}\sqrt{\frac{N+1}{N+2}}\sum\limits_{l=0}^{M-m}(-1)^l {M-m \choose l}\sum\limits_{k,r}2^{\frac{-k-r}{2}}\mathcal{S}^{(M)}_{N+1}z_{N+2}^{r}z_{N+1}^{k}S_{MN+M-m-l-r} S_{MN+l-k}\Psi_{N}^m,\ee
where we finally fix the arbitrary normalization constants via
\be
 \frac{{{\cal N}_{N+2}^m}'}{{{\cal N}_{N}^m}'}(-1)^{M-m}2^{\frac{2MN+M-m}{2}}2\pi\sqrt{\frac{N+2}{N+1}}=1\,.
\ee
\Eq{nplus2} is equivalent to
 \begin{align}\label{equi_form}\Psi_{N+2}^m=\frac{1}{\sqrt{N+2}}\sum\limits_{l=0}^{M-m}&(-1)^l {M-m \choose l}\sum\limits_{k,r}\sqrt{r!k!} \frac{z_{N+2}^{r}}{\sqrt{2\pi 2^r r!}}\sqrt{N+1}\,\mathcal{S}^{(M)}_{N+1} \frac{z_{N+1}^{k}}{\sqrt{2\pi 2^k k!}} \notag\\&\times S_{MN+M-m-l-r}S_{MN+l-k}\Psi_{N}^m.\end{align}
Since we rigorously derived the above to yield the manifestly (anti-)symmetric
wave function \eqref{mwave}, we may {\em optionally} act on it with the (anti-)symmetrizer
$\mathcal{S}^{(M)}_{N+2}$, giving
\begin{align}\label{sym_equi_form}\Psi_{N+2}^m=\frac{1}{N+2}\sum\limits_{l=0}^{M-m}&(-1)^l {M-m \choose l}\sum\limits_{k,r}\sqrt{r!k!}\sqrt{N+2}\,\mathcal{S}^{(M)}_{N+2} \frac{z_{N+2}^{r}}{\sqrt{2\pi 2^r r!}}\sqrt{N+1}\,\mathcal{S}^{(M)}_{N+1} \frac{z_{N+1}^{k}}{\sqrt{2\pi 2^k k!}}\notag\\&\times  S_{MN+M-m-l-r}S_{MN+l-k}\Psi_{N}^m.\end{align}
Upon second quantization  by using Eq.~(1.13) of Ref. \onlinecite{cr2nd}, with \Eq{spwfdisk} in mind, the above formula leads to
\be
\ket{\text{Pf}_{N+2}^m}= \frac{1}{N+2}\sum\limits_{l=0}^{M-m}(-1)^l {M-m \choose l} \sum\limits_{r,k=0}^{MN+M-m }\sqrt{r!k!}\, c_{r}^\dagger c_{k}^\dagger  S_{MN+M-m-l-r}S_{MN+l-k}\ket{\text{Pf}_{N}^m}, \ee of which \Eq{refor1} is a special case with $m=1$.

Now we derive the expression for  general Pfaffian state with one particle removed by using Eq.~(1.12) of Ref. \onlinecite{cr2nd} (Gaussians are included in the integration measure):
\begin{align}\label{anni_nplus2}c_r\Psi_{N+2}^m=&\sqrt{N+2}\int d^2z_{N+2} \frac{\overline{z}_{N+2}^{r}}{\sqrt{2\pi2^r r!}}\Psi_{N+2}^m.\end{align}
We now see why we went through the effort to not only derive Eq. \eqref{sym_equi_form}, which could have been arrived at more directly, but instead took the pains to also derive Eq. \eqref{equi_form}. This equation has the much needed advantage to expose the dependence on
$z_{N+2}$ by having this variable appear outside of the symmetrization.
Via the change of variable $l \rightarrow M-m-l$ we rewrite \Eq{equi_form} as:
\begin{align}\Psi_{N+2}^m=\frac{1}{2\sqrt{N+2}}\sum\limits_{l=0}^{M-m}&(-1)^l {M-m \choose l}\sum\limits_{k,r}\sqrt{r!k!} \frac{z_{N+2}^{r}}{\sqrt{2\pi 2^r r!}}\sqrt{N+1}\,\mathcal{S}^{(M)}_{N+1} \frac{z_{N+1}^{k}}{\sqrt{2\pi 2^k k!}}\notag\\&\times [ S_{MN+M-m-l-r} S_{MN+l-k}+(-1)^{M-m}S_{MN+M-m-l-k}S_{MN+l-r} ]\Psi_{N}^m.\end{align}
Then, \Eq{anni_nplus2} leads to
\begin{align}
c_r \ket{\text{Pf}_{N+2}^m}=\frac{\sqrt{r!}}{2}\sum\limits_{l=0}^{M-m}&(-1)^l {M-m \choose l} \sum\limits_{k=0}^{MN+M-m }\sqrt{k!}\, c_{k}^\dagger \notag\\ &\times (S_{MN+M-m-l-r}S_{MN+l-k}  +(-1)^{M-m}S_{MN+M-m-l-k}S_{MN+l-r})\ket{\text{Pf}_{N}^m},\end{align}
of which \Eq{re1} is a special case with $m=1$.

\section{The annihilation of $\ket{\text{Pf}_{4}}$  by all  $Q^{(2\text{bd},\cal P)}_J$}\label{2annihilate}
By using the recursive formula \Eq{refor1}, the second-quantized form of $\ket{\text{Pf}_{4}}$ is \begin{align}\label{psi_4}\ket{\text{Pf}_{4}}=&\frac{1}{8}\sum\limits_{p_1,p_2,q_1,q_2=0}^{M}(-1)^{\sum_{i=1}^2 (p_i+q_i)} \prod_{i=1}^2{M \choose p_i}{M \choose q_i} \sum\limits_{l_1,l_2=0}^{M-1}(-1)^{l_1+l_2}\prod_{i=1}^2{M-1 \choose l_i}\notag\\&\times \sqrt{(3M-1-l_2-q_1-q_2)! (2M+l_2-p_1-p_2)! (M-1-l_1+p_1+q_1)! (l_1+p_2+q_2)!} \notag\\&\times   c_{3M-1-l_2-q_1-q_2}^\dagger c_{2M+l_2-p_1-p_2}^\dagger   c_{M-1-l_1+p_1+q_1}^\dagger c_{l_1+p_2+q_2}^\dagger  \ket{0},\end{align}
where we have used the commutator \be\label{scdag}[S_{l},c^{\dagger}_{r}]=\sum_{k=1}^{M}(-1)^k {M \choose k} \sqrt{\frac{(r+k)!}{r!}}c_{r+k}^\dagger S_{l-k} \ee to move $S$ to the right of $c^\dagger$. We act with $Q^{(2\text{bd},\cal P)}_J$ on $\ket{\text{Pf}_{4}}$ to obtain
\begin{align}&Q^{(2\text{bd},\cal P)}_J\ket{\text{Pf}_{4}}\notag\\=&\frac{1}{4}\sum\limits_{p_1,p_2,q_1,q_2=0}^{M}(-1)^{p_1+p_2+q_1+q_2} {M \choose p_1}{M \choose p_2}{M \choose q_1} {M \choose q_2}\sum\limits_{l_2=0}^{M-1}(-1)^{l_2}{M-1 \choose l_2} \notag\\&\quad\times\Bigg[\sum\limits_{l_1=0}^{M-1}(-1)^{l_1}{M-1 \choose l_1}  {\cal P}(M-1-l_1+p_1+q_1,l_1+p_2+q_2)\Bigg]\delta_{J,M+p_1+p_2+q_1+q_2-1}\notag\\&\quad\times \sqrt{(3M-1-l_2-q_1-q_2)!(2M+l_2-p_1-p_2)!}\,c_{3M-1-l_2-q_1-q_2}^\dagger c_{2M+l_2-p_1-p_2}^\dagger    \ket{0} \notag\\&+\frac{1}{4}\sum\limits_{p_1,q_1,q_2=0}^{M}(-1)^{p_1+q_1+q_2} {M \choose p_1}{M \choose q_1} {M \choose q_2}\sum\limits_{l_1,l_2=0}^{M-1}(-1)^{l_1+l_2}{M-1 \choose l_1} {M-1 \choose l_2}\notag\\&\quad\times\Bigg[\sum\limits_{p_2=0}^{M}(-1)^{p_2}{M \choose p_2} {\cal P}(l_1+p_2+q_2,2M+l_2-p_1-p_2)\Bigg]\delta_{J,l_1+l_2+2 M-p_1+q_2}\notag\\&\quad\times \sqrt{(3M-1-l_2-q_1-q_2)!(M-1-l_1+p_1+q_1)!}\,c_{3M-1-l_2-q_1-q_2}^\dagger   c_{M-1-l_1+p_1+q_1}^\dagger  \ket{0} \notag\\&+\frac{1}{4}\sum\limits_{p_2,q_1,q_2=0}^{M}(-1)^{p_2+q_1+q_2}{M \choose p_2}{M \choose q_1} {M \choose q_2}\sum\limits_{l_1,l_2=0}^{M-1}(-1)^{l_1+l_2}{M-1 \choose l_1} {M-1 \choose l_2}\notag\\&\quad\times\Bigg[\sum\limits_{p_1=0}^{M}(-1)^{p_1} {M \choose p_1}  {\cal P}(2M+l_2-p_1-p_2 ,M-1-l_1+p_1+q_1 )\Bigg]\delta_{J,-l_1+l_2+3 M-p_2+q_1-1}\notag\\&\quad\times \sqrt{(3M-1-l_2-q_1-q_2)!(l_1+p_2+q_2)!}\,c_{3M-1-l_2-q_1-q_2}^\dagger  c_{l_1+p_2+q_2}^\dagger  \ket{0} \notag\\&+\frac{1}{4}\sum\limits_{p_1,p_2,q_1=0}^{M}(-1)^{p_1+p_2+q_1} {M \choose p_1}{M \choose p_2}{M \choose q_1} \sum\limits_{l_1,l_2=0}^{M-1}(-1)^{l_1+l_2}{M-1 \choose l_1} {M-1 \choose l_2}\notag\\&\quad\times\Bigg[\sum\limits_{q_2=0}^{M}(-1)^{q_2}  {M \choose q_2}  {\cal P}(3M-1-l_2-q_1-q_2 ,l_1+p_2+q_2)\Bigg]\delta_{J,l_1-l_2+3 M+p_2-q_1-1}\notag\\&\quad\times \sqrt{(2M+l_2-p_1-p_2)!(M-1-l_1+p_1+q_1)!}\,c_{2M+l_2-p_1-p_2}^\dagger  c_{M-1-l_1+p_1+q_1}^\dagger   \ket{0} \notag\\&+\frac{1}{4}\sum\limits_{p_1,p_2,q_2=0}^{M}(-1)^{p_1+p_2+q_2} {M \choose p_1}{M \choose p_2} {M \choose q_2}\sum\limits_{l_1,l_2=0}^{M-1}(-1)^{l_1+l_2}{M-1 \choose l_1} {M-1 \choose l_2}\notag\\&\quad\times\Bigg[\sum\limits_{q_1=0}^{M}(-1)^{q_1} {M \choose q_1}  {\cal P}(M-1-l_1+p_1+q_1 ,3M-1-l_2-q_1-q_2)\Bigg]\delta_{J,-l_1-l_2+4 M+p_1-q_2-2}\notag\\&\quad\times \sqrt{(2M+l_2-p_1-p_2)!(l_1+p_2+q_2)!}\,c_{2M+l_2-p_1-p_2}^\dagger  c_{l_1+p_2+q_2}^\dagger  \ket{0} \notag\\&+\frac{1}{4}\sum\limits_{p_1,p_2,q_1,q_2=0}^{M}(-1)^{p_1+p_2+q_1+q_2} {M \choose p_1}{M \choose p_2}{M \choose q_1} {M \choose q_2}\sum\limits_{l_1=0}^{M-1}(-1)^{l_1}{M-1 \choose l_1} \notag\\&\quad\times\Bigg[\sum\limits_{l_2=0}^{M-1}(-1)^{l_2}{M-1 \choose l_2} {\cal P}(3M-1-l_2-q_1-q_2,2M+l_2-p_1-p_2)\Bigg]\delta_{J,5 M-p_1-p_2-q_1-q_2-1}\notag\\&\quad\times \sqrt{(M-1-l_1+p_1+q_1)!(l_1+p_2+q_2)!}\,c_{M-1-l_1+p_1+q_1}^\dagger c_{l_1+p_2+q_2}^\dagger  \ket{0} \notag\\=&0,\end{align} where the summation in each of $4 \choose 2$  square  brackets is zero by using a combinatorial identity\cite{Ruiz96} \be\label{combi} \sum_{i=0}^{n} (-1)^{i} {n \choose i}i^p=0 \quad \text{for any integer} \ p \in [0,n-1], \ee  considering that the degree of $\cal P$ is less than $M-1$.

\section{The annihilation of $\ket{\text{Pf}_{4}}$  and $\ket{\text{Pf}_{6}}$ by all $Q^{(3\text{bd},\cal Q)}_J$}\label{3annihilate}
The second-quantized form of $\ket{\text{Pf}_{4}}$ has been given in \Eq{psi_4}, and the second-quantized form of $\ket{\text{Pf}_{6}}$ is \begin{align}\ket{\text{Pf}_{6}}=&  \frac{1}{48}\sum\limits_{l_1,l_2,l_3=0}^{M-1}(-1)^{\sum_{i=1}^3 l_i}\prod_{i=1}^3{M-1 \choose l_i} \sum\limits_{p_1,\cdots,p_6,q_1,\cdots,q_6=0}^{M}(-1)^{\sum_{i=1}^6 (p_i+q_i)} \prod_{i=1}^6{M \choose p_i}{M \choose q_i}  \notag\\
&\times     \sqrt{(5M-1-l_3-q_3-q_4-q_5-q_6)! (4M+l_3-p_3-p_4-p_5-p_6)!(3M-1-l_2-q_1-q_2+p_3+q_3)!}  \notag\\&\times \sqrt{ (2M+l_2-p_1-p_2+p_4+q_4)! (M-1-l_1+p_1+q_1+p_5+q_5)!(l_1+p_2+q_2+p_6+q_6)!}\notag\\&\times c_{5M-1-l_3-q_3-q_4-q_5-q_6}^\dagger     c_{4M+l_3-p_3-p_4-p_5-p_6}^\dagger    c_{3M-1-l_2-q_1-q_2+p_3+q_3}^\dagger c_{2M+l_2-p_1-p_2+p_4+q_4}^\dagger  c_{M-1-l_1+p_1+q_1+p_5+q_5}^\dagger \notag\\&\times    c_{l_1+p_2+q_2+p_6+q_6}^\dagger   \ket{0}. \end{align}
We act $Q^{(3\text{bd},\cal Q)}_J$ on $\ket{\text{Pf}_{4}}$ to obtain
\begin{align}\label{3bodyon4}&Q^{(3\text{bd},\cal Q)}_J\ket{\text{Pf}_{4}}\notag\\
=&\frac{3(-1)^{M-1}}{4}\sum\limits_{q_1,q_2=0}^{M}(-1)^{q_1+q_2} {M \choose q_1} {M \choose q_2}\sum\limits_{l_2=0}^{M-1}(-1)^{l_2}{M-1 \choose l_2}\notag\\&\quad\times\Bigg[\sum\limits_{p_1,p_2=0}^{M}\sum\limits_{l_1=0}^{M-1}(-1)^{l_1}{M-1 \choose l_1} (-1)^{p_1+p_2} {M \choose p_1}{M \choose p_2}\notag\\&\quad\times{\cal Q}(2M+l_2-p_1-p_2,M-1-l_1+p_1+q_1,l_1+p_2+q_2)
\Bigg]\notag\\&\quad\times \delta_{J,l_2+3 M+q_1+q_2-1}\sqrt{(3M-1-l_2-q_1-q_2)!}    \, c_{3M-1-l_2-q_1-q_2}^\dagger   \ket{0} \notag\\
&+\frac{3}{4}\sum\limits_{p_1,p_2=0}^{M}(-1)^{p_1+p_2} {M \choose p_1}{M \choose p_2}\sum\limits_{l_2=0}^{M-1}(-1)^{l_2}{M-1 \choose l_2} \notag\\&\quad\times\Bigg[\sum\limits_{q_1,q_2=0}^{M}\sum\limits_{l_1=0}^{M-1}(-1)^{l_1}{M-1 \choose l_1}(-1)^{q_1+q_2} {M \choose q_1} {M \choose q_2}\notag\\&\quad\times{\cal Q}(3M-1-l_2-q_1-q_2,M-1-l_1+p_1+q_1,l_1+p_2+q_2)\Bigg] \notag\\&\quad\times  \delta_{J,-l_2+4 M+p_1+p_2-2}\sqrt{(2M+l_2-p_1-p_2)!}  \,  c_{2M+l_2-p_1-p_2}^\dagger   \ket{0} \notag\\
&+\frac{3(-1)^{M-1}}{4}\sum\limits_{p_1,q_1=0}^{M}(-1)^{p_1+q_1} {M \choose p_1}{M \choose q_1} \sum\limits_{l_1=0}^{M-1}(-1)^{l_1}{M-1 \choose l_1} \notag\\&\quad\times\Bigg[\sum\limits_{p_2,q_2=0}^{M}\sum\limits_{l_2=0}^{M-1}(-1)^{l_2}{M-1 \choose l_2}(-1)^{p_2+q_2}{M \choose p_2} {M \choose q_2}\notag\\&\quad\times{\cal Q}(3M-1-l_2-q_1-q_2,2M+l_2-p_1-p_2,l_1+p_2+q_2)\Bigg] \notag\\&\quad\times \delta_{J,l_1+5 M-p_1-q_1-1}\sqrt{(M-1-l_1+p_1+q_1)!}   \,   c_{M-1-l_1+p_1+q_1}^\dagger   \ket{0} \notag\\
&+\frac{3}{4}\sum\limits_{p_2,q_2=0}^{M}(-1)^{p_2+q_2}{M \choose p_2} {M \choose q_2}\sum\limits_{l_1=0}^{M-1}(-1)^{l_1}{M-1 \choose l_1}  \notag\\&\quad\times\Bigg[\sum\limits_{p_1,q_1=0}^{M}\sum\limits_{l_2=0}^{M-1}(-1)^{l_2}{M-1 \choose l_2}(-1)^{p_1+q_1} {M \choose p_1}{M \choose q_1}\notag\\&\quad\times{\cal Q}(3M-1-l_2-q_1-q_2,2M+l_2-p_1-p_2,M-1-l_1+p_1+q_1)\Bigg]\notag\\&\quad\times \delta_{J,-l_1+6 M-p_2-q_2-2} \sqrt{(l_1+p_2+q_2)!}    \,  c_{l_1+p_2+q_2}^\dagger  \ket{0} \notag\\=&0,\end{align} where the term in each of $4 \choose 3$ square brackets is zero.
Take the first square bracket as an example: on account of \Eq{combi}, for the summations inside the first square bracket not to vanish, there should exist at least one term in $\cal Q$ in which the power of $l_1$, $p_1$  and  $p_2$ should be greater than or equal to $M-1$,   $M$ and  $M$, respectively. However, the degree of  $\cal Q$ is less than $3M-1$.  Therefore, the term in the first square bracket vanishes. Likewise, summations in all other square brackets are zero.

Along the same line of logic, it is easy to verify $Q^{(3\text{bd},\cal Q)}_J\ket{\text{Pf}_{6}}=0$.

\end{widetext}
\bibliography{Pf}

\widetext
\newpage
\begin{center}
\textbf{\large Supplemental Material for ``From frustration-free parent Hamiltonians to off-diagonal long-range order: Moore-Read and related states in second quantization"}
\end{center}

\setcounter{equation}{0}
\setcounter{figure}{0}
\setcounter{table}{0}
\setcounter{page}{1}
\setcounter{section}{0}
\makeatletter
\renewcommand{\theequation}{S\arabic{equation}}
\renewcommand{\thefigure}{S\arabic{figure}}
\renewcommand{\bibnumfmt}[1]{[S#1]}
\renewcommand{\citenumfont}[1]{S#1}

In this  supplemental material, our aim is to prove by mathematical induction the formula
\begin{align}\label{reformula}
c_r \ket{\text{Pf}_{N+2}}=\frac{\sqrt{r!}}{2}\sum\limits_{l=0}^{M-m}(-1)^l {M-m \choose l} \sum\limits_{k=0}^{MN+M-m }&\sqrt{k!}\, c_{k}^\dagger  (S_{MN+M-m-l-r}S_{MN+l-k} \notag\\ & +(-1)^{M-m}S_{MN+M-m-l-k}S_{MN+l-r})\ket{\text{Pf}_{N}},\end{align}
with  even/odd positive integer $M$ for fermionic/bosonic case and odd positive integer $m$ obeying $1\leqslant m \leqslant M$, where $\ket{\text{Pf}_{N}}$ and $\ket{\text{Pf}_{N+2}}$  are 
states of $N$ and $N+2$ particles with even $N$, related by the recursive formula
 \be\label{generalre}
\ket{\text{Pf}_{N+2}}= \frac{1}{N+2}\sum\limits_{l=0}^{M-m}(-1)^l {M-m \choose l} \sum\limits_{r,k=0}^{MN+M-m }\sqrt{r!k!}\, c_{r}^\dagger c_{k}^\dagger  S_{MN+M-m-l-r}S_{MN+l-k}\ket{\text{Pf}_{N}}.  \ee
In first quantization, the states so defined correspond to the generalized Pfaffian
wave function

\be\text{Pf}\left[\frac{1}{(z_i-z_j)^m}\right]\prod\limits_{k<l}\left(z_k-z_l\right)^M,\ee
with the standard Moore-Read state corresponding to the special case $m=1$. It is important, however,
that we will not use this first quantized expression in the following. This completes the reasoning of the main text that all known properties of the parent Hamiltonian of the Moore-Read state can be inferred in second quantization.


Let us study the beginning of mathematical induction.
We have $\ket{\text{Pf}_{0}}=\ket{0}$, and
\begin{align}
\ket{\text{Pf}_{2}}=&  \frac{1}{2}\sum\limits_{l=0}^{M-m}(-1)^l {M-m \choose l} \sum\limits_{r,k=0}^{M-m }\sqrt{r!k!}\, c_{r}^\dagger c_{k}^\dagger  S_{M-m-l-r}S_{l-k}\ket{0}\notag\\=&  \frac{1}{2}\sum\limits_{l=0}^{M-m}(-1)^l {M-m \choose l} \sqrt{(M-m-l)!l!}\, c_{M-m-l}^\dagger c_{l}^\dagger \ket{0}.\end{align}
In the calculation of $\ket{\text{Pf}_{2}}$, we have used the fact that  the $S$ operator is the sum of products of $e$ operators, which have annihilation operators on the right, thus $S_{M-m-l-r}S_{l-k}\ket{0}$ gives zero unless $M-m-l-r=0$ and $l-k=0$. From $\ket{\text{Pf}_{0}}$ and $\ket{\text{Pf}_{2}}$, we have an identity
\be\begin{split}c_r \ket{\text{Pf}_{2}}=\frac{\sqrt{r!}}{2}\sum\limits_{l=0}^{M-m}(-1)^l {M-m \choose l} \sum\limits_{k=0}^{M-m} &\sqrt{k!}\, c_{k}^\dagger  (S_{M-m-l-r}S_{l-k}  +(-1)^{M-m}S_{M-m-l-k}S_{l-r})\ket{\text{Pf}_{0}}\end{split}  \ee manifestly satisfied for all $r$, as seen from acting $c_r $ on $\ket{\text{Pf}_{2}}$.

Now we assume \be\label{reformulaN-2}\begin{split}
c_r \ket{\text{Pf}_{N}}=\frac{\sqrt{r!}}{2}\sum\limits_{l=0}^{M-m}(-1)^l {M-m \choose l} \sum\limits_{k=0}^{MN-M-m }&\sqrt{k!}\, c_{k}^\dagger  (S_{MN-M-m-l -r}S_{MN-2M+l -k} \\ & +(-1)^{M-m}S_{MN-M-m-l -k}S_{MN-2M+l -r})\ket{\text{Pf}_{N-2}},\end{split}  \ee is valid for even $N\geqslant 2$,
we then have \begin{subequations}\begin{align}
c_r \ket{\text{Pf}_{N+2}} =&c_r \frac{1}{N+2}\sum\limits_{l=0}^{M-m}(-1)^l {M-m \choose l} \sum\limits_{j,k=0}^{MN+M-m }\sqrt{j!k!}\, c_{j}^\dagger c_{k}^\dagger  S_{MN+M-m-l -j}S_{MN+l-k}\ket{\text{Pf}_{N}}\notag\\ =&\frac{2}{N+2}\frac{\sqrt{r!}}{2}\sum\limits_{l=0}^{M-m}(-1)^l {M-m \choose l}\sum\limits_{k=0}^{MN+M-m }\sqrt{k!}\,  c_{k}^\dagger  \notag\\ &\times (S_{MN+M-m-l-r}S_{MN+l-k}  +(-1)^{M-m}S_{MN+M-m-l-k}S_{MN+l-r})\ket{\text{Pf}_{N}} \label{1}\\ &+\frac{1}{N+2}\sum\limits_{l=0}^{M-m}(-1)^l {M-m \choose l}\sum\limits_{j,k=0}^{MN+M-m }\sqrt{j!k!}\, c_{j}^\dagger c_{k}^\dagger  c_{r} S_{MN+M-m-l -j}S_{MN+l-k} \ket{\text{Pf}_{N}}\label{2}. \end{align}\end{subequations}
Using the commutator \be[c_{i}, S_{l}]=\sum_{k=1}^{M} (-1)^k {M \choose  k} \sqrt{\frac{i!}{(i-k)!}}S_{l-k}c_{i-k}\ee  
(and, as usual, the convention $c_{i-k}=0$ for $k>i$),
\Eq{2} can be written as \be\label{2b} \begin{split}
&\frac{1}{N+2}\sum\limits_{l=0}^{M-m}(-1)^l {M-m \choose l} \sum\limits_{j,k=0}^{MN+M-m }\sqrt{j!k!}\, c_{j}^\dagger c_{k}^\dagger  \sum\limits_{q_1,q_2=0}^{M} (-1)^{q_1+q_2}{M \choose  q_1}{M \choose  q_2}\\&\times  \sqrt{\frac{r!}{(r-q_1-q_2)!}} S_{MN+M-m-l -j-q_1}S_{MN+l-k-q_2}c_{r-q_1-q_2}\ket{\text{Pf}_{N}}.\end{split}  \ee
By using the induction assumption \Eq{reformulaN-2},  \Eq{2b}  can be further cast as \be\label{2c}\begin{split}
&\frac{1/2}{N+2}  \sum\limits_{l=0}^{M-m}(-1)^l {M-m \choose l} \sum\limits_{l'=0}^{M-m}(-1)^{l'} {M-m \choose l'}\sum\limits_{j,k=0}^{MN+M-m }\sum\limits_{p=0}^{MN-M-m }\sqrt{j!k!p!r!}\\&\times \sum\limits_{q_1,q_2=0}^{M} (-1)^{q_1+q_2}{M \choose  q_1}{M \choose  q_2} c_{j}^\dagger c_{k}^\dagger  S_{MN+M-m-l -j-q_1}S_{MN+l-k-q_2}c_{p}^\dagger \\ &\times (S_{MN-M-m-l' -r+q_1+q_2}S_{MN-2M+l' -p}   +(-1)^{M-m}S_{MN-M-m-l' -p}S_{MN-2M+l' -r+q_1+q_2})\ket{\text{Pf}_{N-2}}.\end{split}  \ee Now we  split  \Eq{2c} into the $q_1=q_2=M $ part
\be\label{2prime}\begin{split}
&\frac{1/2}{N+2}  \sum\limits_{l=0}^{M-m}(-1)^l {M-m \choose l} \sum\limits_{l'=0}^{M-m}(-1)^{l'} {M-m \choose l'}\sum\limits_{j,k=0}^{MN+M-m }\sum\limits_{p=0}^{MN-M-m }\sqrt{j!k!p!r!}\\&\times  c_{j}^\dagger c_{k}^\dagger  S_{MN-m-l -j}S_{MN-M+l-k}c_{p}^\dagger  \\ &\times(S_{MN+M-m-l' -r}S_{MN-2M+l' -p}  +(-1)^{M-m}S_{MN-M-m-l' -p}S_{MN+l' -r})\ket{\text{Pf}_{N-2}}.\end{split}  \ee
and the rest that we denote as
\be\begin{split}
G_1:=&\frac{1/2}{N+2}  \sum\limits_{l=0}^{M-m}(-1)^l {M-m \choose l} \sum\limits_{l'=0}^{M-m}(-1)^{l'} {M-m \choose l'}\left(\sum\limits_{q_1=0}^{M-1} \sum\limits_{q_2=0}^{M-1} +\sum\limits_{q_1=0}^{M-1} \sum\limits_{q_2=M }^{M} +\sum\limits_{q_1=M }^{M} \sum\limits_{q_2=0}^{M-1} \right)\\&\times \sum\limits_{j,k=0}^{MN+M-m }\sum\limits_{p=0}^{MN-M-m }\sqrt{j!k!p!r!}(-1)^{q_1+q_2}{M \choose  q_1}{M \choose  q_2} c_{j}^\dagger c_{k}^\dagger  S_{MN+M-m-l -j-q_1}S_{MN+l-k-q_2}c_{p}^\dagger   \\&\times (S_{MN-M-m-l' -r+q_1+q_2}S_{MN-2M+l' -p}  +(-1)^{M-m}S_{MN-M-m-l' -p}S_{MN-2M+l' -r+q_1+q_2})\ket{\text{Pf}_{N-2}}.\end{split}  \ee
Now in \Eq{2prime}, we move $c_{p}^\dagger$ to the left of $S_{MN-m-l -j}S_{MN-M+l -k}$  using the commutator \be\label{scd}[S_{l},c^{\dagger}_{r}]=\sum_{q=1}^{M} (-1)^q {M \choose  q} \sqrt{\frac{(r+q)!}{r!}}c_{r+q}^\dagger S_{l-q},\ee and regard everything other than the $q=M $ term as $G_2$ and $G_3$:

\Eq{2prime}
\be\label{2pp}\begin{split}
=&\frac{1/2}{N+2}  \sum\limits_{l=0}^{M-m}(-1)^l {M-m \choose l} \sum\limits_{l'=0}^{M-m}(-1)^{l'} {M-m \choose l'}\sum\limits_{j,k=0}^{MN+M-m }\sum\limits_{p=0}^{MN-M-m }\sqrt{j!k!(p+2M)!r!}\,c_{j}^\dagger c_{k}^\dagger  c_{p+2M}^\dagger\\&\times    S_{MN-M-m-l -j}S_{MN-2M+l-k} (S_{MN+M-m-l' -r}S_{MN-2M+l' -p}  +(-1)^{M-m}S_{MN-M-m-l' -p}S_{MN+l' -r})\ket{\text{Pf}_{N-2}}.\end{split}  \ee
plus
\be\begin{split}
G_2:=&\frac{1/2}{N+2}  \sum\limits_{l=0}^{M-m}(-1)^l {M-m \choose l} \sum\limits_{l'=0}^{M-m}(-1)^{l'} {M-m \choose l'}\sum\limits_{j,k=0}^{MN+M-m }\sum\limits_{p=0}^{MN-M-m }\sqrt{j!k!(p+q)!r!}
\\&\times \sum_{q=0}^{M-1} (-1)^q {M \choose  q}c_{j}^\dagger c_{k}^\dagger S_{MN-m-l -j}c_{p+q}^\dagger S_{MN-M+l-k-q} \\&\times(S_{MN+M-m-l' -r}S_{MN-2M+l' -p}  +(-1)^{M-m}S_{MN-M-m-l' -p}S_{MN+l' -r})\ket{\text{Pf}_{N-2}},\end{split}  \ee
plus
\be\begin{split}
G_3:=&\frac{(-1)^{M}/2}{N+2}  \sum\limits_{l=0}^{M-m}(-1)^l {M-m \choose l} \sum\limits_{l'=0}^{M-m}(-1)^{l'} {M-m \choose l'}\sum\limits_{j,k=0}^{MN+M-m }\sum\limits_{p=0}^{MN-M-m }\sqrt{j!k!(p+M+q)!r!}  \\&\times \sum_{q=0}^{M-1} (-1)^q {M \choose  q}c_{j}^\dagger c_{k}^\dagger  c_{p+M+q}^\dagger S_{MN-m-l -j-q}S_{MN-2M+l-k}  \\&\times(S_{MN+M-m-l' -r}S_{MN-2M+l' -p}  +(-1)^{M-m}S_{MN-M-m-l' -p}S_{MN+l' -r})\ket{\text{Pf}_{N-2}}.\end{split}  \ee
In \Eq{2pp},
we have every right
to extend the $p$-sum to negative
values,
since for such negative $p$ values, both $MN-2M+l' -p $ and $MN-M-m-l' -p$ are larger than  $M(N-2)$, therefore
$S_{MN-2M+l' -p}$
and $S_{MN-M-m-l' -p}$
annihilate
the  $(N-2)$-particle state $\ket{\text{Pf}_{N-2}}$. The reason  is the following:  $S_i$  is expressed as $( - 1)^i \sum_{{n_1} + {n_2}+\cdots+{n_M}= i}  e_{n_1} e_{n_2}\cdots e_{n_M}$, in which $e_{n}$ will move the orbitals of $n$ particles. For  $i>M(N-2)$, there must be an $n$  larger than  $N-2$, thus $S_i$ annihilates  $\ket{\text{Pf}_{N-2}}$ in this case. We thus let the $p$-sum start at $-2M$,
and then let $p \rightarrow p- 2M$.  This gives

\Eq{2pp}
\begin{align}\label{2ppp}
=&\frac{1/2}{N+2}  \sum\limits_{l=0}^{M-m}(-1)^l {M-m \choose l} \sum\limits_{l'=0}^{M-m}(-1)^{l'} {M-m \choose l'}\sum\limits_{j,k,p=0}^{MN+M-m }\sqrt{j!k!p!r!}\, c_{j}^\dagger c_{k}^\dagger  c_{p}^\dagger S_{MN-M-m-l -j}S_{MN-2M+l-k} \notag\\&\times    (S_{MN+M-m-l' -r}S_{MN+l' -p} +(-1)^{M-m}S_{MN+M-m-l' -p}S_{MN+l' -r})\ket{\text{Pf}_{N-2}} \notag\\
=&\frac{1/2}{N+2}  \sum\limits_{l=0}^{M-m}(-1)^l {M-m \choose l} \sum\limits_{l'=0}^{M-m}(-1)^{l'} {M-m \choose l'}\sum\limits_{j,k,p=0}^{MN+M-m }\sqrt{j!k!p!r!}\, c_{p}^\dagger c_{j}^\dagger c_{k}^\dagger   \notag\\&\times    (S_{MN+l' -p}S_{MN+M-m-l' -r}  +(-1)^{M-m}S_{MN+M-m-l' -p}S_{MN+l' -r}) S_{MN-M-m-l -j} S_{MN-2M+l-k} \ket{\text{Pf}_{N-2}},\end{align}
where we have used the commutability of $S$ operators and the commutability of $c_{j}^\dagger c_{k}^\dagger$ with $c_{p}^\dagger$.

Finally, we move $c_{j}^\dagger c_{k}^\dagger $ in \Eq{2ppp} to the right of  $(S_{MN+M-m-l'-r}S_{MN+l'-p}  +(-1)^{M-m}S_{MN+M-m-l'-p}S_{MN+l'-r})$, use the variation of  \Eq{scd} \be  c^{\dagger}_{r}S_{l} =S_{l} c^{\dagger}_{r}+\sum_{q=1}^{M} (-1)^{q+1} {M \choose  q} \sqrt{\frac{(r+q)!}{r!}}c_{r+q}^\dagger S_{l-q},\ee and regard everything other than the $q=0$ term as $G_4$, $G_5$, $G_6$, and $G_7$:

\Eq{2ppp}
\begin{align}\label{2pppp}
=&\frac{1/2}{N+2}  \sum\limits_{l=0}^{M-m}(-1)^l {M-m \choose l} \sum\limits_{l'=0}^{M-m}(-1)^{l'} {M-m \choose l'}\sum\limits_{j,k,p=0}^{MN+M-m }\sqrt{j!k!p!r!}\, c_{p}^\dagger  \notag\\ & \times(S_{MN+M-m-l' -r}S_{MN+l' -p}  +(-1)^{M-m}S_{MN+M-m-l' -p}S_{MN+l' -r})   c_{j}^\dagger c_{k}^\dagger  \notag\\ & \times S_{MN-M-m-l -j} S_{MN-2M+l-k} \ket{\text{Pf}_{N-2}}\end{align}
plus
\begin{align}
G_4:=&\frac{1/2}{N+2}  \sum\limits_{l=0}^{M-m}(-1)^l {M-m \choose l} \sum\limits_{l'=0}^{M-m}(-1)^{l'} {M-m \choose l'}  \sum\limits_{j,k,p=0}^{MN+M-m }\sqrt{j!(k+q)!p!r!} \sum_{q=1}^{M} (-1)^{q+1} {M \choose  q}\notag\\&\times c_{p}^\dagger c_{j}^\dagger c_{k+q}^\dagger
 (S_{MN+l'-p-q}S_{MN+M-m-l'-r}  +(-1)^{M-m}S_{MN+M-m-l'-p-q}S_{MN+l'-r}) \notag\\&\times S_{MN-M-m-l -j} S_{MN-2M+l -k}\ket{\text{Pf}_{N-2}},\end{align}
plus
\begin{align}
G_5:=&\frac{1/2}{N+2}  \sum\limits_{l=0}^{M-m}(-1)^l {M-m \choose l} \sum\limits_{l'=0}^{M-m}(-1)^{l'} {M-m \choose l'}  \sum\limits_{j,k,p=0}^{MN+M-m }\sqrt{j!(k+q)!p!r!}\sum\limits_{q=1}^{M} (-1)^{q+1}{M \choose  q} \notag\\&\times c_{p}^\dagger c_{j}^\dagger
    (S_{MN+l'-p}c_{k+q}^\dagger S_{MN+M-m-l'-r-q}+(-1)^{M-m}S_{MN+M-m-l'-p}c_{k+q}^\dagger S_{MN+l'-r-q}) \notag\\&\times S_{MN-M-m-l -j}S_{MN-2M+l -k}\ket{\text{Pf}_{N-2}},\end{align}
plus
\begin{align}
G_6:=&\frac{1/2}{N+2}  \sum\limits_{l=0}^{M-m}(-1)^l {M-m \choose l} \sum\limits_{l'=0}^{M-m}(-1)^{l'} {M-m \choose l'}  \sum\limits_{j,k,p=0}^{MN+M-m }\sqrt{(j+q)!k!p!r!}\sum\limits_{q=1}^{M} (-1)^{q+1}{M \choose  q}\notag\\&\times c_{p}^\dagger c_{j+q}^\dagger
    (S_{MN+l'-p-q}S_{MN+M-m-l'-r}+(-1)^{M-m}S_{MN+M-m-l'-p-q}S_{MN+l'-r})c_{k}^\dagger\notag\\&\times
    S_{MN-M-m-l -j}S_{MN-2M+l -k}\ket{\text{Pf}_{N-2}},\end{align}
plus
\begin{align}
G_7:=&\frac{1/2}{N+2}  \sum\limits_{l=0}^{M-m}(-1)^l {M-m \choose l} \sum\limits_{l'=0}^{M-m}(-1)^{l'} {M-m \choose l'}  \sum\limits_{j,k,p=0}^{MN+M-m }\sqrt{(j+q)!k!p!r!}\sum\limits_{q=1}^{M} (-1)^{q+1}{M \choose  q}\notag\\&\times c_{p}^\dagger
    (S_{MN+l'-p}c_{j+q}^\dagger S_{MN+M-m-l'-r-q}+(-1)^{M-m}S_{MN+M-m-l'-p}c_{j+q}^\dagger S_{MN+l'-r-q}) c_{k}^\dagger\notag\\&\times  S_{MN-M-m-l -j}S_{MN-2M+l -k}\ket{\text{Pf}_{N-2}}.\end{align}
Now \Eq{2pppp} can be written as
\begin{align}
&\frac{1/2}{N+2}   \sum\limits_{l'=0}^{M-m}(-1)^{l'} {M-m \choose l'}\sum\limits_{p=0}^{MN+M-m }\sqrt{p!r!}\,c_{p}^\dagger     (S_{MN+M-m-l' -r}S_{MN+l' -p} +(-1)^{M-m}S_{MN+M-m-l' -p}S_{MN+l' -r})     \notag\\ & \times \sum\limits_{l=0}^{M-m}(-1)^l {M-m \choose l}\sum\limits_{j,k=0}^{MN-M-m }\sqrt{j!k!}\,c_{j}^\dagger c_{k}^\dagger S_{MN-M-m-l -j} S_{MN-2M+l-k} \ket{\text{Pf}_{N-2}}\label{15a}\\=&\frac{1/2}{N+2}   \sum\limits_{l'=0}^{M-m}(-1)^{l'} {M-m \choose l'}\sum\limits_{p=0}^{MN+M-m }\sqrt{p!r!}\,c_{p}^\dagger     (S_{MN+M-m-l' -r}S_{MN+l' -p} +(-1)^{M-m}S_{MN+M-m-l' -p}S_{MN+l' -r})     \notag\\ & \times N \ket{\text{Pf}_{N}}\label{15b}.\end{align}
From \Eq{2pppp} to \Eq{15a}, we have changed the summation range of both $j$ and  $k$ by using the property that  $S_{i}=0$ for negative $i$. \Eq{15a} leads to \Eq{15b} by using the recursive formula from $\ket{\text{Pf}_{N-2}}$ to $\ket{\text{Pf}_{N}}$.

Combining all the above terms, we obtain \begin{align}\label{result}
c_r \ket{\text{Pf}_{N+2}} =&\frac{\sqrt{r!}}{2}\sum\limits_{l=0}^{M-m}(-1)^l {M-m \choose l}   \sum\limits_{k=0}^{MN+M-m }\sqrt{k!}\,  c_{k}^\dagger  \notag\\ &\times (S_{MN+M-m-l-r}S_{MN+l-k}   +(-1)^{M-m}S_{MN+M-m-l-k}S_{MN+l-r})\ket{\text{Pf}_{N}}\notag\\ &+\sum\limits_{i=1}^{7} G_i. \end{align}

It is easiest to compare these $G_i$ terms
after commuting all $c^\dagger$-operators to the
left. This will produce terms with one, two, three, and four $q$-sums:
\begin{align}
G_1=&\frac{1/2}{N+2}  \sum\limits_{l=0}^{M-m}(-1)^l {M-m \choose l} \sum\limits_{l'=0}^{M-m}(-1)^{l'} {M-m \choose l'}  \sum\limits_{j,k=0}^{MN+M-m }\sum\limits_{p=0}^{MN-M-m }\left(\sum\limits_{q_1=0}^{M-1} \sum\limits_{q_2=0}^{M-1} +\sum\limits_{q_1=0}^{M-1} \sum\limits_{q_2=M }^{M} +\sum\limits_{q_1=M }^{M} \sum\limits_{q_2=0}^{M-1} \right) \notag\\&\times \sum _{q_3=0}^{M}  \sum _{q_4=0}^{M}  \sqrt{ j!k!  (p+q_3+q_4)!r!} (-1)^{q_1+q_2+q_3+q_4}{M \choose  q_1}{M \choose  q_2}{M \choose  q_3}{M \choose  q_4}\notag\\&\times  c_{j}^\dagger c_{k}^\dagger  c_{p+q_3+q_4}^\dagger  S_{MN+M-m-l -j-q_1-q_4}S_{MN+l-k-q_2-q_3} \notag\\ & \times(S_{MN-M-m-l' -r+q_1+q_2}S_{MN-2M+l' -p}  +(-1)^{M-m}S_{MN-M-m-l' -p}S_{MN-2M+l' -r+q_1+q_2})\ket{\text{Pf}_{N-2}},\end{align}

\begin{align}
G_2 =&\frac{1/2}{N+2}  \sum\limits_{l=0}^{M-m}(-1)^l {M-m \choose l} \sum\limits_{l'=0}^{M-m}(-1)^{l'} {M-m \choose l'}  \sum\limits_{j,k=0}^{MN+M-m }\sum\limits_{p=0}^{MN-M-m }\sum _{q_1=0}^{M-1}  \sum _{q_2=0}^{M}  \sqrt{ j!k!  (p+q_1+q_2)!r!}\notag\\&\times   (-1)^{q_1+q_2}{M \choose  q_1}{M \choose  q_2}c_{j}^\dagger c_{k}^\dagger c_{p+q_1+q_2}^\dagger S_{MN-m-l -j-q_2} S_{MN-M+l-k-q_1} \notag\\ & \times(S_{MN+M-m-l' -r}S_{MN-2M+l' -p}   +(-1)^{M-m}S_{MN-M-m-l' -p}S_{MN+l' -r})\ket{\text{Pf}_{N-2}},\end{align}

\begin{align}
G_3=&\frac{(-1)^{M}/2}{N+2}  \sum\limits_{l=0}^{M-m}(-1)^l {M-m \choose l} \sum\limits_{l'=0}^{M-m}(-1)^{l'} {M-m \choose l'}  \sum\limits_{j,k=0}^{MN+M-m }\sum\limits_{p=0}^{MN-M-m }\sum _{q=0}^{M-1}  \sqrt{ j!k!  (p+q+M)!r!}(-1)^{q}{M \choose  q}\notag\\&\times c_{j}^\dagger c_{k}^\dagger  c_{p+q+M}^\dagger S_{MN-m-l -j-q}S_{MN-2M+l-k} \notag\\ & \times (S_{MN+M-m-l' -r}S_{MN-2M+l' -p}   +(-1)^{M-m}S_{MN-M-m-l' -p}S_{MN+l' -r})\ket{\text{Pf}_{N-2}},\end{align}

\begin{align}
G_4 =&\frac{(-1)^M/2}{N+2}  \sum\limits_{l=0}^{M-m}(-1)^l {M-m \choose l} \sum\limits_{l'=0}^{M-m}(-1)^{l'} {M-m \choose l'}  \sum\limits_{j,k,p=0}^{MN+M-m }\sum _{q=1}^{M}  \sqrt{ j!k!  (p+q)!r!}(-1)^{q}{M \choose  q}c_{j}^\dagger c_{k}^\dagger  c_{p+q}^\dagger\notag\\&\times   S_{MN-M-m-l -j} S_{MN-2M+l -p} \notag\\&\times  (S_{MN+l'-k-q}S_{MN+M-m-l'-r}  +(-1)^{M-m}S_{MN+M-m-l'-k-q}S_{MN+l'-r})\ket{\text{Pf}_{N-2}},\end{align}
(where we have made change of variables $k\leftrightarrow p$ in $G_4$)

\begin{align}
G_5 =&\frac{(-1)^M/2}{N+2}  \sum\limits_{l=0}^{M-m}(-1)^l {M-m \choose l} \sum\limits_{l'=0}^{M-m}(-1)^{l'} {M-m \choose l'}  \sum\limits_{j,k,p=0}^{MN+M-m }\sum _{q_1=1}^{M}  \sum _{q_2=0}^{M}  \sqrt{ j!k!  (p+q_1+q_2)!r!}(-1)^{q_1+q_2}\notag\\&\times{M \choose  q_1}{M \choose  q_2}  c_{j}^\dagger  c_{k}^\dagger c_{p+q_1+q_2}^\dagger S_{MN-M-m-l -j}S_{MN-2M+l -p}  \notag\\ & \times (S_{MN+l'-k-q_2} S_{MN+M-m-l'-r-q_1}  +(-1)^{M-m}S_{MN+M-m-l'-k-q_2} S_{MN+l'-r-q_1})  \ket{\text{Pf}_{N-2}},\end{align}
(where we have made change of variables $k\leftrightarrow p$ in $G_5$)

\begin{align}
G_6 =&-\frac{1/2}{N+2}  \sum\limits_{l=0}^{M-m}(-1)^l {M-m \choose l} \sum\limits_{l'=0}^{M-m}(-1)^{l'} {M-m \choose l'}  \sum\limits_{j,k,p=0}^{MN+M-m }\sum _{q_1=1}^{M}  \sum _{q_2=0}^{M}  \sum _{q_3=0}^{M}   \sqrt{ (j+q_1)! (k+q_2+q_3)!p!r!}\notag\\&\times   (-1)^{q_1+q_2+q_3}{M \choose  q_1}{M \choose  q_2}{M \choose  q_3}   c_{j+q_1}^\dagger c_{k+q_2+q_3}^\dagger c_{p}^\dagger  S_{MN-M-m-l -j}S_{MN-2M+l -k}\notag\\&\times(S_{MN+l'-p-q_1-q_3}S_{MN+M-m-l'-r-q_2}+(-1)^{M-m}S_{MN+M-m-l'-p-q_1-q_3}S_{MN+l'-r-q_2})\ket{\text{Pf}_{N-2}},\end{align}

and \begin{align}
G_7=&-\frac{1/2}{N+2}  \sum\limits_{l=0}^{M-m}(-1)^l {M-m \choose l} \sum\limits_{l'=0}^{M-m}(-1)^{l'} {M-m \choose l'}  \sum\limits_{j,k,p=0}^{MN+M-m }\sum _{q_1=1}^{M}  \sum _{q_2=0}^{M}  \sum _{q_3=0}^{M}  \sum _{q_4=0}^{M}  \notag\\&\times\sqrt{ (j+q_1+q_2)!(k+q_3+q_4)! p!r!}  \, (-1)^{q_1+q_2+q_3+q_4}{M \choose  q_1}{M \choose  q_2}{M \choose  q_3}{M \choose  q_4}\notag\\&\times    c_{j+q_1+q_2}^\dagger  c_{k+q_3+q_4}^\dagger  c_{p}^\dagger
  S_{MN-M-m-l -j}S_{MN-2M+l -k} \notag\\&\times   (S_{MN+l'-p-q_2-q_4} S_{MN+M-m-l'-r-q_1-q_3}+(-1)^{M-m}S_{MN+M-m-l'-p-q_2-q_4}S_{MN+l'-r-q_1-q_3})   \ket{\text{Pf}_{N-2}}.\end{align}
We can now apply change of variables to all $G_i$. Take  $G_7$ as an example,  we may  let  $j \rightarrow j-q_1-q_2$ and $k \rightarrow  k-q_3-q_4$.
We may restore the beginnings of these sums to $j=0$ and  $k=0$ as we did from \Eq{2pp} to \Eq{2ppp}. This gives \begin{align}
G_7 =&-\frac{1/2}{N+2}  \sum\limits_{l=0}^{M-m}(-1)^l {M-m \choose l} \sum\limits_{l'=0}^{M-m}(-1)^{l'} {M-m \choose l'}  \sum\limits_{j=0}^{MN+M-m +q_1+q_2}\sum\limits_{k=0}^{MN+M-m +q_3+q_4}\sum\limits_{p=0}^{MN+M-m }\notag\\&\times\sum _{q_1=1}^{M}  \sum _{q_2=0}^{M}  \sum _{q_3=0}^{M}  \sum _{q_4=0}^{M}    \sqrt{  j!k!p! r!}\,  (-1)^{q_1+q_2+q_3+q_4}{M \choose  q_1}{M \choose  q_2}{M \choose  q_3}{M \choose  q_4}  \notag\\&\times  c_{j}^\dagger c_{k}^\dagger c_p^\dagger   S_{MN-M-m-l -j+q_1+q_2}S_{MN-2M+l -k+q_3+q_4} \notag\\&\times   (S_{MN+l'-p-q_2-q_4} S_{MN+M-m-l'-r-q_1-q_3}+(-1)^{M-m}S_{MN+M-m-l'-p-q_2-q_4}S_{MN+l'-r-q_1-q_3})   \ket{\text{Pf}_{N-2}}.\end{align}

We may restore the upper boundaries of both $j$ and $k$ to $MN+M-m$ since $S_{MN-M-m-l -j+q_1+q_2}=0$ for $j > MN+M-m$ and $S_{MN-2M+l -k+q_3+q_4}=0$ for $k >MN+M-m$.

After change of variables,  we obtain new forms for all $G_i$:
\begin{align}
G_1 =&\frac{1/2}{N+2}  \sum\limits_{l=0}^{M-m}(-1)^l {M-m \choose l} \sum\limits_{l'=0}^{M-m}(-1)^{l'} {M-m \choose l'}  \sum\limits_{j,k,p=0}^{MN+M-m } \sqrt{ j!k!  p!r!}\,c_{j}^\dagger c_{k}^\dagger  c_{p}^\dagger  \notag\\&\times \left(\sum\limits_{q_1=0}^{M-1} \sum\limits_{q_2=0}^{M-1} +\sum\limits_{q_1=0}^{M-1} \sum\limits_{q_2=M }^{M} +\sum\limits_{q_1=M }^{M} \sum\limits_{q_2=0}^{M-1} \right) \sum _{q_3=0}^{M}  \sum _{q_4=0}^{M} (-1)^{q_1+q_2+q_3+q_4}{M \choose  q_1}{M \choose  q_2}{M \choose  q_3}{M \choose  q_4}\notag\\&\times   S_{MN+M-m-l -j-q_1-q_4}S_{MN+l-k-q_2-q_3}\notag\\&\times (S_{MN-M-m-l' -r+q_1+q_2}S_{MN-2M+l' -p+q_3+q_4}   +(-1)^{M-m}S_{MN-M-m-l' -p+q_3+q_4}S_{MN-2M+l' -r+q_1+q_2})\ket{\text{Pf}_{N-2}},\end{align}

\begin{align}
G_2=&\frac{1/2}{N+2}  \sum\limits_{l=0}^{M-m}(-1)^l {M-m \choose l} \sum\limits_{l'=0}^{M-m}(-1)^{l'} {M-m \choose l'}  \sum\limits_{j,k,p=0}^{MN+M-m } \sqrt{ j!k! p!r!}\,c_{j}^\dagger c_{k}^\dagger c_{p}^\dagger  \notag\\&\times  \sum _{q_1=0}^{M-1}  \sum _{q_2=0}^{M} (-1)^{q_1+q_2}{M \choose  q_1}{M \choose  q_2} S_{MN-m-l -j-q_2} S_{MN-M+l-k-q_1}\notag\\&\times  (S_{MN+M-m-l' -r}S_{MN-2M+l' -p+q_1+q_2}   +(-1)^{M-m}S_{MN-M-m-l' -p+q_1+q_2}S_{MN+l' -r})\ket{\text{Pf}_{N-2}},\end{align}

\begin{align}
G_3 =&\frac{(-1)^M/2}{N+2}  \sum\limits_{l=0}^{M-m}(-1)^l {M-m \choose l} \sum\limits_{l'=0}^{M-m}(-1)^{l'} {M-m \choose l'}  \sum\limits_{j,k,p=0}^{MN+M-m } \sqrt{ j!k! p!r!}\,c_{j}^\dagger c_{k}^\dagger  c_{p}^\dagger \notag\\&\times  \sum _{q=0}^{M-1} (-1)^{q}{M \choose  q}S_{MN-m-l -j-q}S_{MN-2M+l-k} \notag\\&\times   (S_{MN+M-m-l' -r}S_{MN-M+l' -p+q}    +(-1)^{M-m}S_{MN-m-l' -p+q}S_{MN+l' -r})\ket{\text{Pf}_{N-2}},\end{align}

\begin{align}
G_4=&\frac{(-1)^M/2}{N+2}  \sum\limits_{l=0}^{M-m}(-1)^l {M-m \choose l} \sum\limits_{l'=0}^{M-m}(-1)^{l'} {M-m \choose l'}  \sum\limits_{j,k,p=0}^{MN+M-m }\sqrt{ j!k! p!r!}\,c_{j}^\dagger c_{k}^\dagger  c_{p}^\dagger \notag\\&\times   \sum _{q=1}^{M}  (-1)^{q}{M \choose  q} S_{MN-M-m-l -j} S_{MN-2M+l -p+q}\notag\\&\times (S_{MN+l'-k-q}S_{MN+M-m-l'-r}  +(-1)^{M-m}S_{MN+M-m-l'-k-q}S_{MN+l'-r}) \ket{\text{Pf}_{N-2}},\end{align}

\begin{align}
G_5 =&\frac{(-1)^M/2}{N+2}  \sum\limits_{l=0}^{M-m}(-1)^l {M-m \choose l} \sum\limits_{l'=0}^{M-m}(-1)^{l'} {M-m \choose l'}  \sum\limits_{j,k,p=0}^{MN+M-m }\sqrt{ j!k! p!r!}\,c_{j}^\dagger  c_{k}^\dagger c_{p}^\dagger \notag\\&\times \sum _{q_1=1}^{M}  \sum _{q_2=0}^{M}  (-1)^{q_1+q_2}{M \choose  q_1}{M \choose  q_2}   S_{MN-M-m-l -j}S_{MN-2M+l -p+q_1+q_2}  \notag\\&\times (S_{MN+l'-k-q_2} S_{MN+M-m-l'-r-q_1}  +(-1)^{M-m}S_{MN+M-m-l'-k-q_2} S_{MN+l'-r-q_1})  \ket{\text{Pf}_{N-2}},\end{align}

\begin{align}
G_6=&-\frac{1/2}{N+2}  \sum\limits_{l=0}^{M-m}(-1)^l {M-m \choose l} \sum\limits_{l'=0}^{M-m}(-1)^{l'} {M-m \choose l'}  \sum\limits_{j,k,p=0}^{MN+M-m }\sqrt{j! k!p!r!} c_{j}^\dagger c_{k}^\dagger c_{p}^\dagger    \notag\\&\times  \sum _{q_1=1}^{M}  \sum _{q_2=0}^{M}  \sum _{q_3=0}^{M} (-1)^{q_1+q_2+q_3}{M \choose  q_1}{M \choose  q_2}{M \choose  q_3}   S_{MN-M-m-l -j+q_1}S_{MN-2M+l -k+q_2+q_3}\notag\\&\times(S_{MN+M-m-l'-r-q_2}S_{MN+l'-p-q_1-q_3}+(-1)^{M-m}S_{MN+M-m-l'-p-q_1-q_3}S_{MN+l'-r-q_2})\ket{\text{Pf}_{N-2}},\end{align}

and \begin{align}
G_7=&-\frac{1/2}{N+2}  \sum\limits_{l=0}^{M-m}(-1)^l {M-m \choose l} \sum\limits_{l'=0}^{M-m}(-1)^{l'} {M-m \choose l'}  \sum\limits_{j,k,p=0}^{MN+M-m }\sqrt{  j!k!p! r!}\,c_{j}^\dagger c_{k}^\dagger c_p^\dagger    \notag\\&\times  \sum _{q_1=1}^{M}  \sum _{q_2=0}^{M}  \sum _{q_3=0}^{M}  \sum _{q_4=0}^{M}  (-1)^{q_1+q_2+q_3+q_4}{M \choose  q_1}{M \choose  q_2}{M \choose  q_3}{M \choose  q_4}  \notag\\&\times   S_{MN-M-m-l -j+q_1+q_2}S_{MN-2M+l -k+q_3+q_4} \notag\\&\times   (S_{MN+l'-p-q_2-q_4} S_{MN+M-m-l'-r-q_1-q_3}+(-1)^{M-m}S_{MN+M-m-l'-p-q_2-q_4}S_{MN+l'-r-q_1-q_3})   \ket{\text{Pf}_{N-2}}.\end{align}

Now  we apply change of variables $q_i=M-q_i$ for $i=1,2,3,4$ to $G_1$:
\begin{align}
G_1=&\frac{1/2}{N+2}  \sum\limits_{l=0}^{M-m}(-1)^l {M-m \choose l} \sum\limits_{l'=0}^{M-m}(-1)^{l'} {M-m \choose l'}  \sum\limits_{j,k,p=0}^{MN+M-m } \sqrt{ j!k!  p!r!}\,c_{j}^\dagger c_{k}^\dagger  c_{p}^\dagger  \notag\\&\times \left(\sum\limits_{q_1=1}^{M} \sum\limits_{q_2=1}^{M} +\sum\limits_{q_1=1}^{M} \sum\limits_{q_2=0}^{0} +\sum\limits_{q_1=0}^{0} \sum\limits_{q_2=1}^{M} \right) \sum _{q_3=0}^{M}  \sum _{q_4=0}^{M} (-1)^{q_1+q_2+q_3+q_4}{M \choose  q_1}{M \choose  q_2}{M \choose  q_3}{M \choose  q_4}\notag\\&\times   S_{MN-M-m-l -j+q_1+q_4}S_{MN-2M+l-k+q_2+q_3}\notag\\&\times (S_{MN+M-m-l' -r-q_1-q_2}S_{MN+l' -p-q_3-q_4}   +(-1)^{M-m}S_{MN+M-m-l' -p-q_3-q_4}S_{MN+l' -r-q_1-q_2})\ket{\text{Pf}_{N-2}},\end{align}
We also relabel $q_2 \rightarrow q_4$, $q_3\rightarrow q_2$ and $q_4 \rightarrow q_3$ in $G_7$:
\begin{align}
G_7=&-\frac{1/2}{N+2}  \sum\limits_{l=0}^{M-m}(-1)^l {M-m \choose l} \sum\limits_{l'=0}^{M-m}(-1)^{l'} {M-m \choose l'}  \sum\limits_{j,k,p=0}^{MN+M-m }\sqrt{  j!k!p! r!}\,c_{j}^\dagger c_{k}^\dagger c_p^\dagger    \notag\\&\times  \sum _{q_1=1}^{M}  \sum _{q_2=0}^{M}  \sum _{q_3=0}^{M}  \sum _{q_4=0}^{M}  (-1)^{q_1+q_2+q_3+q_4}{M \choose  q_1}{M \choose  q_2}{M \choose  q_3}{M \choose  q_4}  \notag\\&\times   S_{MN-M-m-l -j+q_1+q_4}S_{MN-2M+l -k+q_2+q_3} \notag\\&\times   (S_{MN+M-m-l'-r-q_1-q_2}S_{MN+l'-p-q_4-q_3} +(-1)^{M-m}S_{MN+M-m-l'-p-q_4-q_3}S_{MN+l'-r-q_1-q_2})   \ket{\text{Pf}_{N-2}}.\end{align}
Combining $G_1$ and $G_7$, we obtain a new expression  \begin{align}
G'_1: =&\frac{1/2}{N+2}  \sum\limits_{l=0}^{M-m}(-1)^l {M-m \choose l} \sum\limits_{l'=0}^{M-m}(-1)^{l'} {M-m \choose l'}  \sum\limits_{j,k,p=0}^{MN+M-m } \sqrt{ j!k!  p!r!}\,c_{j}^\dagger c_{k}^\dagger  c_{p}^\dagger  \notag\\&\times  \sum\limits_{q_2=1}^{M} \sum _{q_3=0}^{M}  \sum _{q_4=0}^{M} (-1)^{q_2+q_3+q_4}{M \choose  q_2}{M \choose  q_3}{M \choose  q_4}\notag\\&\times   S_{MN-M-m-l -j+q_4}S_{MN-2M+l-k+q_2+q_3}\notag\\&\times (S_{MN+M-m-l' -r-q_2}S_{MN+l' -p-q_3-q_4}   +(-1)^{M-m}S_{MN+M-m-l' -p-q_3-q_4}S_{MN+l' -r-q_2})\ket{\text{Pf}_{N-2}},\end{align}
We then relabel $q_4 \rightarrow q_1$  in $G'_1$, add up $G'_1$ and $G_6$ to obtain $G''_1$:
\begin{align}
G''_1: =&\frac{1/2}{N+2}  \sum\limits_{l=0}^{M-m}(-1)^l {M-m \choose l} \sum\limits_{l'=0}^{M-m}(-1)^{l'} {M-m \choose l'}  \sum\limits_{j,k,p=0}^{MN+M-m }\sqrt{j! k!p!r!} c_{j}^\dagger c_{k}^\dagger c_{p}^\dagger    \notag\\&\times  \sum _{q_2=0}^{M}  \sum _{q_3=0}^{M} (-1)^{q_2+q_3}{M \choose  q_2}{M \choose  q_3}   S_{MN-M-m-l -j}S_{MN-2M+l -k+q_2+q_3}\notag\\&\times(S_{MN+M-m-l'-r-q_2}S_{MN+l'-p-q_3}+(-1)^{M-m}S_{MN+M-m-l'-p-q_3}S_{MN+l'-r-q_2})\ket{\text{Pf}_{N-2}}\notag\\ &-\frac{1/2}{N+2}  \sum\limits_{l=0}^{M-m}(-1)^l {M-m \choose l} \sum\limits_{l'=0}^{M-m}(-1)^{l'} {M-m \choose l'}  \sum\limits_{j,k,p=0}^{MN+M-m }\sqrt{j! k!p!r!} c_{j}^\dagger c_{k}^\dagger c_{p}^\dagger    \notag\\&\times  \sum _{q_1=0}^{M}    \sum _{q_3=0}^{M} (-1)^{q_1+q_3}{M \choose  q_1}{M \choose  q_3}   S_{MN-M-m-l -j+q_1}S_{MN-2M+l -k+q_3}\notag\\&\times(S_{MN+M-m-l'-r}S_{MN+l'-p-q_1-q_3}+(-1)^{M-m}S_{MN+M-m-l'-p-q_1-q_3}S_{MN+l'-r})\ket{\text{Pf}_{N-2}},\end{align}
Using this method, we add up $G''_1$,  $G_2$ and $G_5$ to obtain $G'''_1$:
\begin{align}
G'''_1:=&-\frac{(-1)^M/2}{N+2}  \sum\limits_{l=0}^{M-m}(-1)^l {M-m \choose l} \sum\limits_{l'=0}^{M-m}(-1)^{l'} {M-m \choose l'}  \sum\limits_{j,k,p=0}^{MN+M-m }\sqrt{ j!k! p!r!}\,c_{j}^\dagger  c_{k}^\dagger c_{p}^\dagger \notag\\&\times  \sum _{q_2=0}^{M}  (-1)^{q_2}{M \choose  q_2}   S_{MN-M-m-l -j}S_{MN-2M+l -p+q_2}  \notag\\&\times (S_{MN+l'-k-q_2} S_{MN+M-m-l'-r}  +(-1)^{M-m}S_{MN+M-m-l'-k-q_2} S_{MN+l'-r})  \ket{\text{Pf}_{N-2}}\notag\\ &-\frac{1/2}{N+2}  \sum\limits_{l=0}^{M-m}(-1)^l {M-m \choose l} \sum\limits_{l'=0}^{M-m}(-1)^{l'} {M-m \choose l'}  \sum\limits_{j,k,p=0}^{MN+M-m }\sqrt{j! k!p!r!} c_{j}^\dagger c_{k}^\dagger c_{p}^\dagger    \notag\\&\times   \sum _{q_2=0}^{M}    (-1)^{q_2}{M \choose  q_2}   S_{MN-M-m-l -j+q_2}S_{MN-2M+l -k}\notag\\&\times(S_{MN+M-m-l'-r}S_{MN+l'-p-q_2}+(-1)^{M-m}S_{MN+M-m-l'-p-q_2}S_{MN+l'-r})\ket{\text{Pf}_{N-2}},\end{align}
Finally, we obtain
\begin{align}
\sum\limits_{i=1}^{7} G_i=&G'''_1+G_3+G_4\notag\\ =&-\frac{(-1)^M/2}{N+2}  \sum\limits_{l=0}^{M-m}(-1)^l {M-m \choose l} \sum\limits_{l'=0}^{M-m}(-1)^{l'} {M-m \choose l'}  \sum\limits_{j,k,p=0}^{MN+M-m }\sqrt{ j!k! p!r!}\,c_{j}^\dagger  c_{k}^\dagger c_{p}^\dagger \notag\\&\times  \sum _{q=0}^{0}  (-1)^{q}{M \choose  q}   S_{MN-M-m-l -j}S_{MN-2M+l -p+q}  \notag\\&\times (S_{MN+l'-k-q} S_{MN+M-m-l'-r}  +(-1)^{M-m}S_{MN+M-m-l'-k-q} S_{MN+l'-r})  \ket{\text{Pf}_{N-2}}\notag\\ &-\frac{1/2}{N+2}  \sum\limits_{l=0}^{M-m}(-1)^l {M-m \choose l} \sum\limits_{l'=0}^{M-m}(-1)^{l'} {M-m \choose l'}  \sum\limits_{j,k,p=0}^{MN+M-m }\sqrt{j! k!p!r!} c_{j}^\dagger c_{k}^\dagger c_{p}^\dagger    \notag\\&\times   \sum _{q=0}^{0}    (-1)^{q}{M \choose  q}   S_{MN-M-m-l -j+q}S_{MN-2M+l -k}\notag\\&\times(S_{MN+M-m-l'-r}S_{MN+l'-p-q}+(-1)^{M-m}S_{MN+M-m-l'-p-q}S_{MN+l'-r})\ket{\text{Pf}_{N-2}}\notag\\ =&-\frac{(-1)^M/2}{N+2}  \sum\limits_{l=0}^{M-m}(-1)^l {M-m \choose l} \sum\limits_{l'=0}^{M-m}(-1)^{l'} {M-m \choose l'}  \sum\limits_{j,k,p=0}^{MN+M-m }\sqrt{ j!k! p!r!}\,c_{j}^\dagger  c_{k}^\dagger c_{p}^\dagger \notag\\&\times    S_{MN-M-m-l -j}S_{MN-2M+l -p}  \notag\\&\times (S_{MN+l'-k} S_{MN+M-m-l'-r}  +(-1)^{M-m}S_{MN+M-m-l'-k} S_{MN+l'-r})  \ket{\text{Pf}_{N-2}}\notag\\ &-\frac{1/2}{N+2}  \sum\limits_{l=0}^{M-m}(-1)^l {M-m \choose l} \sum\limits_{l'=0}^{M-m}(-1)^{l'} {M-m \choose l'}  \sum\limits_{j,k,p=0}^{MN+M-m }\sqrt{j! k!p!r!} c_{j}^\dagger c_{k}^\dagger c_{p}^\dagger    \notag\\&\times   S_{MN-M-m-l -j}S_{MN-2M+l -k}\notag\\&\times(S_{MN+M-m-l'-r}S_{MN+l'-p}+(-1)^{M-m}S_{MN+M-m-l'-p}S_{MN+l'-r})\ket{\text{Pf}_{N-2}}\notag\\ =&0,\end{align}
where in the last step, we have applied change of variables $k\leftrightarrow p$,  commutability of $S$ operators, and $c_{k}^\dagger c_p^\dagger=(-1)^{M-m}c_p^\dagger c_{k}^\dagger$.

%

From \Eq{result}, we obtain \begin{align}
c_r \ket{\text{Pf}_{N+2}} =\frac{\sqrt{r!}}{2}\sum\limits_{l=0}^{M-m}(-1)^l {M-m \choose l} \sum\limits_{k=0}^{MN+M-m }&\sqrt{k!}\,  c_{k}^\dagger  (S_{MN+M-m-l-r}S_{MN+l-k} \notag\\ &  +(-1)^{M-m}S_{MN+M-m-l-k}S_{MN+l-r})\ket{\text{Pf}_{N}}, \end{align} thus completing our second-quantized derivation.

\end{document}